\documentclass{aims}
\usepackage{paralist}
\usepackage[colorlinks=true]{hyperref}
\hypersetup{urlcolor=blue, citecolor=red}
\usepackage{amsmath,amssymb}
\usepackage{amsfonts}
\usepackage{eucal}
\title[Hamilton--Jacobi theory]{THE HAMILTON--JACOBI THEORY AND THE ANALOGY BETWEEN CLASSICAL AND QUANTUM MECHANICS}

\author[G. Marmo G. Morandi and N. Mukunda]{}

\subjclass{70H20, 70S05, 70FXX, 70HXX}
 \keywords{Hamilton--Jacobi equation, Lagrangian system, Hamiltonian system, Semiclassical Approximation}

\email{marmo@na.infn.it}
\email{morandi@bo.infn.it}
\email{nmukunda@cts.iisc.ernet.in}

\def\<#1>{\langle#1\rangle}

\theoremstyle{definition}

\newtheorem{remark}{Remark}

\def\beq{\begin{equation}}
\def\eeq{\end{equation}}
\def\bea{\begin{eqnarray}}
\def\eea{\end{eqnarray}}
\def\beann{\begin{eqnarray*}}
\def\eeann{\end{eqnarray*}}
\def\ben{\begin{enumerate}}
\def\een{\end{enumerate}}
\def\proof{{\sl Proof}\quad}
\def\qed{\ifvmode\removelastskip\fi
{\unskip\nobreak\hfil\penalty50\hbox{}\nobreak\hfil \hbox{\vrule
height1.2ex width1.2ex}\parfillskip=0pt \finalhyphendemerits=0
\par\smallskip}}

\def\texthook{\vrule height 0pt depth 0.4pt width 3.5pt
          \vrule height 5pt depth 0.4pt \kern 3pt}
\def\scripthook{\vrule height 0pt depth 0.2pt width 1.5pt
                \vrule height 3pt depth 0.2pt width 0.2pt \kern 1pt}

\begin{document}

\maketitle

\centerline{\scshape G. Marmo }
\medskip
{\footnotesize
 \centerline{Dipartimento di Scienze Fisiche and INFN.}
   \centerline{Universit\`a di Napoli "Federico II". v. Cintia. I-80126 Naples, Italy.}
} 

\medskip

\centerline{\scshape G. Morandi and N. Mukunda}
\medskip
{\footnotesize
 \centerline{Dipartimento di Fisica and INFN.}
   \centerline{Universit\`a di Bologna. 6/2 v.le B.Pichat. I-40127 Bologna, Italy.}
   \centerline{Centre for Theoretical Studies.}
   \centerline{Indian Institute of Science. Bangalore 560 012, India.}
}

\bigskip

 \centerline{(Communicated by the associate editor name)}

\begin{abstract}
We review here some conventional as well as less conventional aspects of the time-independent and time-dependent Hamilton-Jacobi ($HJ$) theory and of its connections with Quantum Mechanics. Less conventional aspects involve the $HJ$ theory on the tangent bundle of a configuration manifold, the quantum $HJ$ theory, $HJ$ problems for general differential operators and the $HJ$ problem for Lie groups.
\end{abstract}



\clearpage
\tableofcontents
\clearpage

\section{Introduction}\label{Intro}
\subsection{A Recollection of Preliminary Notions.}
The Hamilton-Jacobi formulation of Classical Dynamics is usually presented
\cite{Am,Ar,Ben,Be,Car,CS,Go,La,MSSV,Ru,SM,Whi2} as the search, for a Hamiltonian
system with a possibly time-dependent Hamiltonian on the cotangent bundle
$T^{\ast}Q$ of some configuration manifold $Q$ (with $\dim Q=n$ for some $n$),
for a canonical transformation that is able to "reduce the system to
equilibrium".The Hamilton-Jacobi $\left(  HJ\right)  $ equation for the
generator $S$ of the transformation, also known as "Hamilton's principal
function", is well known to be then:%
\begin{equation}
H\left(  q;\frac{\partial S}{\partial q};t\right)  +\frac{\partial S}{\partial
t}=0 \label{HJ1}%
\end{equation}
where: $H=H\left(  q;p;t\right)  $ is the Hamiltonian of the system and, if a
\textit{complete integral} $S=S\left(  q;Q\right)  $, i.e. a solution
depending on as many additional parameters $Q^{1},...,Q^{n}$ as \ the number
of degrees of freedom in an essential way, i.e. such that:%
\begin{equation}
\det\left\vert \frac{\partial^{2}S}{\partial q^{i}\partial Q^{j}}\right\vert
\neq0\label{essential}
\end{equation}
is available, then, the canonical transformation: $\left(  q,p\right)
\rightarrow\left(  Q,P\right)  $ defined by:%
\begin{equation}
p_{i}=\frac{\partial S}{\partial q^{i}},\text{ \ }P_{i}=-\frac{\partial
S}{\partial Q^{i}};\text{ }i=1,...n \label{ct1}%
\end{equation}
does the job of reducing the system to equilibrium.

For a time-independent Hamiltonian, and hence for a conservative system,
denoting by $E$ the total energy, the assumption:%
\begin{equation}
S=W-Et \label{characteristics}%
\end{equation}
yields instead the \textit{time-independent} $HJ$ equation:%
\begin{equation}
H\left(  q;\frac{\partial W}{\partial q}\right)  =E \label{HJ2}%
\end{equation}
for "Hamilton's characteristic function" $W$, with the energy entering as one
of the additional parameters on which the principal function has to depend. However, even in this case, the most general solution of Eq.(\ref{HJ1}) need not be linear in $t$.

The $HJ$ equations come from the search for a canonical transformation on the
extended phase space $\mathbb{R}^{2n}\times\mathbb{R}$ such that:%
\begin{equation}
\left(  p_{i}dq^{i}-Hdt\right)  -\left(  P_{i}dQ^{i}-Kdt\right)  =dS\left(
q,Q;t\right)
\end{equation}
with the additional requirement that $K\equiv0$, i.e. that, in the sought-for
new coordinates:%
\begin{equation}
\frac{d}{dt}P_{i}=\frac{d}{dt}Q^{i}=0
\end{equation}
or, more generally:%
\begin{equation}
\frac{d}{dt}f\left(  P,Q;t\right)  \equiv\frac{\partial f}{\partial t}%
\end{equation}
be the "model dynamics" we would like to relate with our starting one, i.e.:%
\begin{equation}
\frac{d}{dt}p_{i}=-\frac{\partial H}{\partial q^{i}};\text{ \ }\frac{dq^{i}%
}{dt}=\frac{\partial H}{\partial p_{i}}%
\end{equation}

Of course, we might consider other "model dynamics". For example, we might
require that: $K=K\left(  P\right)  =\left(  P_{1}^{2}+...+P_{n}^{2}\right)  /2$, in such a way that the
"model dynamics" would be:%
\begin{equation}
\frac{d}{dt}Q^{i}=P_{i};\text{ }\frac{d}{dt}P_{i}=0\text{\ }%
\end{equation}

Clearly, other models could be implemented. For instance, we might require:%
\begin{equation}
\frac{d}{dt}Q^{i}=\nu_{i}P_{i};\text{ }\frac{d}{dt}P_{i}=-\nu_{i}%
Q^{i}\text{\ }%
\end{equation}
and hence: $K=\sum_{i}\nu_{i}(\left(  Q^{i}\right)  ^{2}+\left(  P_{i}\right)
^{2})/2$. The associated partial differential equations would be:
\begin{equation}
H\left(  q,\frac{\partial S}{\partial q};t\right)  +\frac{\partial
S}{\partial
t}=\frac{1}{2}\left(  \frac{\partial S}{\partial Q}\right)  ^{2}%
\end{equation}
or:%
\begin{equation}
H\left(  q,\frac{\partial S}{\partial q};t\right)  +\frac{\partial
S}{\partial
t}=\frac{1}{2}%
{\displaystyle\sum\limits_{j}}
\nu_{j}\left[  \left(  Q^{j}\right)  ^{2}+\left(  \frac{\partial
S}{\partial Q^{j}}\right)  ^{2}\right]
\end{equation}
%

While the first two cases are strictly local, the third one is "less local",
it identifies some equilibrium points, say $Q^{i}=0,P_{i}=0$, which are also
stable ones.

The conditions for solvability of these $PDE$'s are quite strong. For
instance, the first one requires: $dP_{i}/dt=dQ^{i}/dt=0$, i.e. that the
system be maximally integrable. the second one requires the system to be
completely integrable, while the third one requires not only complete
integrability, but also that there be stable equilibrium points.

Thus, the $HJ$ problem in this generalized form would solve not only a
conjugacy problem, i.e. how to transform a given dynamical system into another
one with a preassigned form, but one would find also the required
transformation to be generated by a function $S$.

>From a geometrical point of view, the first case requires the original
dynamics to define a fundamental vector field of the natural foliation of the contact manifold after
we have removed possible equilibrium points. In \ the second case the phase
space is foliated by invariant cylinders, while in the third case the dynamics
will preserve a foliation by tori.

As already noticed, the conditions for the solution of the various problems
are quite stringent, and one should expect that global solutions can be found
only very rarely. Nevertheless, this more general perspective may be
interesting because it could provide the Hamilton-Jacobi theory with a wider
range of applications. For instance, it might be applied to the study of
scattering problems, where now $K$ would be the "comparison Hamiltonian" and
$H$ that of the system one is analyzing. It might be also applicable in Field
Theory and General Relativity. In particular, one might consider applying it
to Quantum Mechanics going beyond the usual $WKBJ$ approximation.

\begin{remark} \textit{Let us observe that if we define, keeping the additional parameters fixed:}%
\begin{equation}
p_{i}=\frac{\partial W}{\partial q^{i}}=p_{i}\left(  q\right)
\end{equation}
\textit{and we form \cite{CGMM} the vector field:}%
\begin{equation}
\mathfrak{X}\left(  Q\right)  \ni X=\frac{\partial H}{\partial p^{i}}\left(
q,p\left(  q\right)  \right)  \frac{\partial}{\partial q^{i}}\label{vf1}%
\end{equation}
\textit{it is immediate to show, using Eq.(\ref{HJ2}), that if: }$q=q\left(
t\right)  $\textit{ is an integral curve for the vector field }$X$\textit{,
then: }$\left(  q\left(  t\right)  ,p\left(  q\left(  t\right)  \right)
\right)  $\textit{ is in turn an integral curve for Hamilton's canonical
equations. Thus, Eq.(\ref{HJ2}) describes in a unified manner one particular family of the phase-space equations of motion.}
\end{remark}
It is worth stressing that, already at this rather well-known level, the
Hamilton-Jacobi theory establishes a deep connection between first-order
partial differential equations $\left(  PDE\text{'s}\right)  $ and systems of
first-order ordinary differential equations $\left(  ODE\text{'s}\right)   $ \cite{Cara}.
\subsection{The $HJ$ Equation and the $JWKB$ Method.}
Quite similar connections appear when one investigates the short-wavelength
limit of "wave-like" equations (including also the Schr\"{o}dinger equation)
such as Hamiltonian optics as the short-wavelength r\'{e}gime of wave optics
(the eikonal approximation \cite{BW,Whi1}, but see also Ref.\cite{LMSV} for
applications in Field Theory) and classical mechanics as the short-wavelength
r\'{e}gime of wave mechanics \cite{Me}.

In taking this limit, one starts with some differential operator of
hyperbolic type on some manifold $Q$, passes through a
Hamilton-Jacobi-type equation (the characteristics equation
\cite{LC}) for a function $S$ and, by substituting covectors (the
$"p"$'s) for the first-order derivatives, arrives at a some
Hamiltonian function on the cotangent bundle $T^{\ast}Q$ \ which
yields also the dispersion relation of the wave motion. The
associated Hamilton equations give rise, by projection on $Q$ of the
solutions, to the bi-characteristics \cite{LC} of the original
differential system. Rephrased in an "optical" language, the
characteristics describe the propagation of wave fronts, while the
bi-characteristics describe that of rays. Note that "substituting
covectors for first-order derivatives", in the case in which the
differential operator is homogeneous, is just a down-to-the -earth
way of saying that we deal with what is known as the
symbol \cite{EMS} of the operator.

Also, the $JWKB$ method \footnote{This method was first introduced in the
discussion of problems in wave propagation by lord Rayleigh back in $1912$ and
then applied to wave mechanics by H.Jeffreys in $1923$ and, later on and
simultaneously, by L.Brillouin, H.A.Kramers and G.Wentzel in $1926$%
.}\ \cite{EMS,Fo,Me,Pe,Sy} is widely used both in wave optics and
wave mechanics to investigate several physical problems in the
short-wavelength r\'{e}gime which, in wave mechanics, amounts to a
leading-term expansion in powers of the Planck constant $\hbar$.
It is also closely related to the saddle-point approximation (plus
one-loop corrections) in the path-integral approach to problems in
Field Theory \cite{Am}, Quantum \cite{GS} and Statistical Mechanics \cite{Kl}.

In this context, and concentrating for the sake of definiteness on the motion
of a particle of mass $m$ in a potential $V\left(  \mathbf{x}\right)  $, one
represents the wave function (in Gaussian form) as\footnote{Many authors \cite{Merz} prefer to incorporate the prefactor $A$ into the
definition of $S$, writing then: $\psi\left(  \mathbf{x.}t\right)
=\exp(iS\left(  \mathbf{x},t\right)  /\hbar$, where the $S$ of Eq.(\ref{WKB}) is
replaced by $S-i\hbar\ln A$ and is no more real.\label{foot}}
:%
\begin{equation}\label{WKB}
\psi\left(  \mathbf{x},t\right)  =A\left(  \mathbf{x},t\right)  \exp\left\{
iS\left(  \mathbf{x},t\right)  /\hbar\right\}
\end{equation}
(with both $A$ and $S$ real). Substituting into the Schr\"{o}dinger equation,
one ends up with the (exact) coupled equations \cite{EMS}:%
\begin{equation}
\frac{1}{2m}\left\vert \nabla S\right\vert ^{2}+V\left(  \mathbf{x}\right)
+\frac{\partial S}{\partial t}=\frac{\hbar^{2}}{2m}\frac{\nabla^{2}A}{A}
\label{HJ3}%
\end{equation}
and:%
\begin{equation}
\nabla\left(  \rho\mathbf{v}\right)  +\frac{\partial\rho}{\partial t}=0
\label{continuity}%
\end{equation}
where: $\rho=\rho\left(  \mathbf{x},t\right)  =A^{2}\left(  \mathbf{x}%
,t\right)  $ and: $\mathbf{v}=\mathbf{v}\left(  \mathbf{x},t\right)  =\nabla
S/m$. Neglecting "quantum correction" proportional to $\hbar^{2}$ on the
r.h.s. of Eq.(\ref{HJ3}) yields the $HJ$ equation (\ref{HJ1}). 
If $S$ solves the time-dependent $HJ$ equation, it is possible to show (see,e.g., App.$4B$ of
Ref.\cite{EMS} and Ref.\cite{Fo}) that Eq.(\ref{continuity}) is
solved by\footnote{The determinant on the r.h.s. of Eq.(\ref{Van
Vleck}) is the well-known \cite{GS,Kl} \textit{Pauli-Morette-Van Vleck
determinant.}}:%
\begin{equation}
\rho=\left\vert \det\left(  \frac{\partial^{2}S}{\partial x^{i}\partial
x_{0}^{j}}\right)  \right\vert \label{Van Vleck}%
\end{equation}
where the $x_{0}^{j}$'s are the initial coordinates, and hence the $JWKB$
solution is:
\begin{equation}
\psi=\sqrt{\rho}\exp\left(  iS/\hbar\right)  \label{propag}%
\end{equation}

As discussed in Ref.\cite{EMS}, Eq.(\ref{propag}) yields the $JWKB$
approximation to the \textit{propagator} (or Green function) for the
Schr\"{o}dinger operator, and the result becomes exact \ for \textit{quadratic
}Hamiltonians. Similar conclusions hold \cite{Am,Kl,LRT} in the path-integral
formalism, where the saddle-point approximation\footnote{In this case the Pauli-Morette-Van
Vleck determinant is replaced by a functional determinant, whose definition
requires careful regularization procedures \cite{GS,Za} that will not be discussed here.}
also becomes exact for quadratic Hamiltonians (or Lagrangians).
\subsection{A Geometrical Setting for the $HJ$ Theory.}
A useful geometrical formulation of the $HJ$ theory can be given as follows.
Consider a symplectic manifold: $\mathcal{M}=T^{\ast}Q$ with symplectic
structure $\omega_{0}=dp_{j}\wedge dq^{j}$. With any symplectomorphism:%
\begin{equation}
\phi:\mathcal{M}\rightarrow\mathcal{M},\text{ \ }\phi^{\ast}\omega_{0}%
=\omega_{0}%
\end{equation}
$\left(  \phi\left(  q,p\right)  =:\left(  Q,P\right)  \right)  $ we may
associate a "graph" $\Sigma_{\phi}$, i.e. a submanifold in the symplectic manifold $\mathcal{M}%
\times\mathcal{M}$, the second factor being equipped with the symplectic
structure \thinspace$-\omega_{0}$:%
\begin{equation}
\Sigma_{\phi}=\left\{  \left(  m,\phi\left(  m\right)  \right)  \subset
\mathcal{M}\times\mathcal{M}\right\}
\end{equation}
by requiring it to be Lagrangian w.r.t. the symplectic structure on
$\mathcal{M}\times\mathcal{M}$ provided by:
\begin{equation}
dp_{j}\wedge dq^{j}-dP_{j}\wedge dQ^{j}=:\omega_{0}\ominus\omega_{0}%
\end{equation}

By using the fact that: $\mathcal{M}\times\mathcal{M}=T^{\ast}Q\times T^{\ast
}Q\rightleftarrows T^{\ast}\left(  Q\times Q\right)  $, we can consider those
Lagrangian submanifolds in $T^{\ast}\left(  Q\times Q\right)  $ that can be
written as graphs:%
\begin{equation}
dS:Q\times Q\longrightarrow T^{\ast}\left(  Q\times Q\right)
\end{equation}
When $\Sigma_{\phi}$ projects onto $Q\times Q$ we may consider it as the graph
of a generating function $S$.

In many cases $\mathcal{M}$ may admit of many alternative cotangent bundle
structures, i.e. we may identify "alternative" submanifolds $Q^{\prime}$ such
that: $\mathcal{M}=T^{\ast}Q^{\prime}$, with $Q$ and $Q^{\prime}$ possible
alternative "placements" of some external configuration space $\mathcal{Q}$. A
typical example is provided by: $\mathcal{M}=\mathbb{R}^{2n}$, where any
identification of \ an affine subspace $\mathbb{R}^{n}$ is \ a possible
placement of an "abstract" $\mathbb{R}^{n}$.

>From this point of view, if we start with $\phi$ and $\Sigma_{\phi}$ we may
look for a particular placement of $Q$ in $\mathcal{M}$ that makes
$\Sigma_{\phi}$ projectable (a submersion onto) $Q\times Q$.

\begin{remark}
The usual "colloquial" classification into "four possible sets of independent
canonical variables" usually encountered in textbooks on Classical Mechanics,
say $\left(  q,Q\right)  ,\left(  q,P\right)  ,\left(  Q,p\right)  $ and
$\left(  p,P\right)  $ are exactly different identifications of the
"configuration space" over which one constructs the cotangent bundle
structure. Therefore we would have the corresponding symplectic structures:
$d\left(  p_{j}dq^{j}-P_{j}dQ^{j}\right)  ,$ $d\left(  p_{j}dq^{j}+Q^{j}%
dP_{j}\right)  ,$ $d\left(  P_{j}dQ^{j}+q^{j}dp_{j}\right)  $ and: $d\left(
q^{j}dp_{j}-Q^{j}dP_{j}\right)  $ respectively.
\end{remark}

When the graph $dS:Q\times Q\rightarrow T^{\ast}\left(  Q\times Q\right)  $ is
viewed as a map: $dS:Q\times Q\rightarrow T^{\ast}Q$, i.e. the second factor
$Q$ is considered as a family of "parameters", we may obtain a regular
foliation of $T^{\ast}Q$. Then we may say that $S$ is a "complete solution" of
some associated $HJ$ equation, and in this case the Hamiltonian flow
associated with the Hamiltonian $K$ would preserve the foliation induced by
$S$ on $T^{\ast}Q$. On the open dense submanifolds on which: $dS:Q\times
Q\rightarrow T^{\ast}Q$ provides a diffeomorphism we would have equivalence
between the Cauchy problem in terms of initial data $\left(  q_{0}%
,p_{0}\right)  $ and the boundary value problem in terms of $\left(
q_{0},Q\right)  $. We notice that in this case:%
\begin{equation}
\left(  dS\right)  ^{\ast}\omega_{0}=:\omega_{S}=\frac{\partial^{2}S}{\partial
q^{i}\partial Q^{j}}dq^{i}\wedge dQ^{J}%
\end{equation}
is a symplectic structure (cfr. Eq.(\ref{essential})) on $Q\times Q$, and
hence it would allow for a Hamiltonian formulation on the space of "boundary
data" rather than of the "initial conditions". The inverse image of the
one-parameter group of evolution on $T^{\ast}Q$ would provide the "propagator"
on the configuration space.

By using the geometrical formulation of Quantum Mechanics a similar picture
can be considered also for Quantum Mechanics in the Schr\"{o}dinger picture.

Let us restrict for convenience to a finite-dimensional Hilbert space
$\mathcal{H}$. Again a transformation:%
\begin{equation}
\Phi:\mathcal{H}\longrightarrow\mathcal{H}%
\end{equation}
will be associated with a graph:%
\begin{equation}
\Sigma_{\Phi}=\left\{  \left(  \psi,\Phi\left(  \psi\right))
\subset \mathcal{H\times H}\right)  \right\}
\end{equation}

If we consider on $\mathcal{H}\times\mathcal{H}$ the pseudo-Hermitian form:%
\begin{equation}
\left\langle \left(  \psi_{1},\psi_{2}\right)  |\left(  \varphi_{1}%
,\varphi_{2}\right)  \right\rangle =:\left\langle \psi_{1}|\varphi
_{1}\right\rangle -\left\langle \psi_{2}|\varphi_{2}\right\rangle
\label{pseudoherm}%
\end{equation}
and noticing that, on $\Sigma_{\Phi}$, $\varphi_{1,2}=\Phi\left(  \psi
_{1,2}\right)  $, we find that \ $\Sigma_{\Phi}$ will be isotropic w.r.t. the
pseudo-Hermitian form (\ref{pseudoherm}) iff $\Phi$ is a unitary transformation.

By considering the realification of $\mathcal{H}$, the Hermitian product
decomposes into a real and an imaginary part, the former providing an
Euclidean product and the latter a symplectic product.

As for the symplectic part, we have \ a situation similar to the one we
considered previously, i.e. we may consider a generating function of the
canonical transformation and require afterwards that this transformation
should preserve also the Euclidean product. In a somewhat simplified notation
we may write then:%
\begin{equation}
\operatorname{Im}\left(  \psi^{\ast}d\psi-\varphi^{\ast}d\varphi\right)
=dS_{\Phi}%
\end{equation}
and then we require, in addition:%
\begin{equation}
\left\langle d\psi|d\psi\right\rangle =\left\langle d\varphi|d\varphi
\right\rangle
\end{equation}

The resulting transformation would be a K\"{a}hlerian transformation derived
from the generating function $S_{\Phi}$. For: $\varphi=U\psi,$ $\varphi^{\ast
}=\psi^{\ast}U^{\ast},$ $S\left(  \varphi^{\ast},\psi\right)  =-\varphi^{\ast
}U\psi$:%
\begin{equation}
d\varphi^{\ast}\wedge d\varphi-d\psi^{\ast}\wedge d\psi=-d\left(  \varphi
d\varphi^{\ast}+\psi^{\ast}d\psi\right)
\end{equation}
For example, a Hadamard gate:%
\begin{equation}
U=U_{H}=\frac{1}{\sqrt{2}}\left\vert
\begin{array}
[c]{cc}%
1 & 1\\
-1 & 1
\end{array}
\right\vert
\end{equation}
has the generating function:%
\begin{equation}
S_{H}\left(  \varphi^{\ast},\psi\right)  =-\varphi_{1}^{\ast}\psi_{1}%
-\varphi_{1}^{\ast}\psi_{2}+\varphi_{2}^{\ast}\psi_{1}-\varphi_{2}^{\ast}%
\psi_{2}%
\end{equation}
\subsection{Quantum $HJ$ Equations.}
In the Heisenberg picture, one may devise a direct approach to the $HJ$
problem by replacing classical variables with Hermitian operators.
Indeed, shortly after the advent of Quantum Mechanics, various efforts were made
\cite{Dir1,Dir3} to formulate both the theory of Canonical Transformations and
hence also the Action Principle \cite{Schw} and the Hamilton-Jacobi equation
in operator terms from the very beginning\footnote{See Ref.\cite{Schw} for
a systematic exposition of this approach.}.

Following, e.g., the scheme of Eqs.(\ref{HJ1}) to (\ref{ct1}), one might be
tempted to "promote" these equations to operator equations defining a
(quantum) canonical transformation as:%
\begin{equation}
\widehat{p}_{i}=\frac{\partial}{\partial\widehat{q}^{i}}S\left(  \widehat
{q},\widehat{Q},t\right)  ,\text{ \ }\widehat{P}_{i}=-\frac{\partial}%
{\partial\widehat{Q}^{i}}S\left(  \widehat{q},\widehat{Q},t\right)
\label{QCT}%
\end{equation}
and a (quantum) $HJ$ equation as:%
\begin{equation}
H\left(  \widehat{q},\frac{\partial S}{\partial\widehat{q}},t\right)
+\frac{\partial}{\partial t}S\left(  \widehat{q},\widehat{Q},t\right)  =0
\label{QHJ}%
\end{equation}

Due to operator-ordering problems, these equations are obviously ambiguous.
According to Jordan \cite{Jo1,Jo2} and Dirac \cite{Dir1,Dir2} the ambiguity should
be resolved by requiring $S$ and the operators in Eqs.(\ref{QCT}) and
(\ref{QHJ}) to be "\textit{well ordered}", i.e. by requiring  all the
"uppercase" operators (the $\widehat{Q}$'s) to stay to the right of the
"lowercase" ones (the $\widehat{q}$'s). This implies that the "generating
operator" $S$ should be of the general form \cite{RS}:%
\begin{equation}
S\left(  \widehat{q},\widehat{Q},t\right)  =\sum\limits_{\alpha}f_{\alpha
}\left(  \widehat{q},t\right)  g_{\alpha}\left(  \widehat{Q},t\right)
\label{QCT2}%
\end{equation}
for suitable functions $f_{\alpha}$ and $g_{\alpha}$. The "well-ordering"
procedure must be applied also, using when necessary the commutation
relations, to the Hamiltonian operator $H$ in Eq.(\ref{QHJ}).

The authors in Ref.\cite{RS} have devised a procedure for converting the
operator equation (\ref{QHJ}) into a $c$-number equation that we will
illustrate here on a simple example\footnote{Referring to Ref.\cite{RS} for
a more general discussion.}, namely that of a (non-relativistic) $1D$ particle
of mass $m$ subject to a scalar potential $V\left(  q\right)  $. The
Hamiltonian is then:%
\begin{equation}
H=\frac{\widehat{p}^{2}}{2m}+V\left(  \widehat{q}\right)
\end{equation}
and the (operator) $HJ$ equation becomes ($S=S(\widehat{q},\widehat{Q},t)$):%
\begin{equation}
\frac{1}{2m}\left(  \frac{\partial S}{\partial\widehat{q}}\right)
^{2}+V\left(  \widehat{q}\right)  +\frac{\partial S}{\partial t}=0
\label{QHJ2}%
\end{equation}

Sandwiching between eigenstates\footnote{$\int dq\left\vert q\rangle\langle
q|\right.  =\mathbb{I}$, $\langle q|q^{\prime}\rangle=\delta\left(
q-q^{\prime}\right)  $, and similarly for $\widehat{Q}$.} $\left\vert
q\rangle\right.  $ and $\left\vert Q\rangle\right.  $ of $\widehat{q}$ and
$\widehat{Q}$ respectively, the "well ordering" prescription leads to:%
\begin{equation}
\langle q|S\left(  \widehat{q},\widehat{Q},t\right)  |Q\rangle=S\left(
q,Q,t\right)  \langle q|Q\rangle
\end{equation}
i.e. to a (uniquely defined and not necessarily real) $c$-number function
$S(q,Q,t)$.

Sandwiching between the same eigenstates the quadratic term on the l.h.s. of
Eq.(\ref{QHJ2}) is a bit more complicated. Explicitly:%
\begin{equation}
\langle q|\left(  \frac{\partial S}{\partial\widehat{q}}\right)  ^{2}%
|Q\rangle=\sum\limits_{\beta}\langle q|\frac{\partial S}{\partial\widehat{q}%
}\frac{\partial f_{\beta}}{\partial\widehat{q}}g_{\beta}(  \widehat
{Q})  |Q\rangle
\end{equation}
Now, the standard canonical commutation relations imply, for any function
$G=G(\widehat{q}):$%
\begin{equation}
\left[  G,\widehat{p}\right]  =i\hbar\frac{\partial G}{\partial \widehat{q}}%
\end{equation}
Using then the first of Eqs.(\ref{QCT}) (i.e.: $\partial S/\partial\widehat
{q}=\widehat{p}$) one finds \cite{RS}:%
\begin{equation}
\frac{\partial S}{\partial\widehat{q}}\frac{\partial f_{\beta}}{\partial
\widehat{q}}=\frac{\partial f_{\beta}}{\partial\widehat{q}}\frac{\partial
S}{\partial\widehat{q}}-i\hbar\frac{\partial^{2}f_{\beta}}{\partial\widehat
{q}^{2}}=\frac{\partial f_{\beta}\left(  \widehat{q},t\right)  }%
{\partial\widehat{q}}\sum\limits_{\alpha}\frac{\partial f_{\alpha}(\widehat
{q},t)}{\partial\widehat{q}}g_{\alpha}(\widehat{Q},t)-i\hbar\frac{\partial
^{2}f_{\beta}(\widehat{q},t)}{\partial\widehat{q}^{2}}%
\end{equation}
which is again a "well ordered" expression, and hence:%
\begin{equation}
\langle q|\left(  \frac{\partial S}{\partial\widehat{q}}\right)  ^{2}%
|Q\rangle=\left[  \left(  \frac{\partial}{\partial q}S\left(  q,Q,t\right)
\right)  ^{2}-i\hbar\frac{\partial^{2}}{\partial q^{2}}S\left(  q,Q,t\right)
\right]  \langle q|Q\rangle
\end{equation}

Dropping then the common factor $\langle q|Q\rangle$ one obtains the
$c$-number equation\footnote{As Eq.(\ref{QHJ3}) (as well as the
classical equation (\ref{HJ1})) contains only derivatives of $S$,
any solution will be ambiguous by the addition of a constant term.
This ambiguity, which is totally irrelevant at the classical level,
will turn out instead
to be useful in what follows. \label{foot2}}:%
\begin{equation}
\frac{1}{2m}\left[  \left(  \frac{\partial}{\partial q}S\left(  q,Q,t\right)
\right)  ^{2}-i\hbar\frac{\partial^{2}}{\partial q^{2}}S\left(  q,Q,t\right)
\right]  +V\left(  q\right)  +\frac{\partial}{\partial t}S\left(
q,Q,t\right)  =0\label{QHJ3}%
\end{equation}
which is completely equivalent to the operator equation (\ref{QHJ2}).

Eq.(\ref{QHJ3}) is precisely the equation that results \cite{Go,RS} from the
Schr\"{o}dinger equation (in the variables $\left(  q,t\right)  $) by
expressing the wave function $\psi$ as (see footnote $\ref{foot}$):%
\begin{equation}
\psi\left(  q,Q,t\right)  =\exp\left\{  \frac{i}{\hbar}S\left(  q,Q,t\right)
\right\}  \label{QHJ4}%
\end{equation}
It provides then a solution of the Schr\"{o}dinger equation depending (in an
essential way) from the additional parameter $Q$, just as the propagator
$K\left(  q,Q,t\right)  $  \cite{EMS,Schu}, which obeys the same equation,
does. The propagator  is known \cite{EMS,Kl,Schu} to obey the boundary condition: $K\left(
q,Q,t\right)  \rightarrow\delta\left(  q-Q\right)  $ as $t\rightarrow0$ and,
in order to complete the identification, one has to check (the equation being
first-order in time) that there is a solution of the form (\ref{QHJ4}) which does the same.
For example, from the well-known\footnote{%
As for any quadratic Hamiltonian.} result \cite{Kl,Schu} for the propagator
of the $1D$ harmonic oscillator with mass $m$, proper frequency $\omega $
and Hamiltonian:%
\begin{equation}
H=\frac{\widehat{p}^{2}}{2m}+\frac{1}{2}m\omega ^{2}\widehat{q}^{2}
\end{equation}
one can check directly that:
\begin{equation}
S=\frac{m\omega }{2\sin \left( \omega t\right) }\left[ \left(
q^{2}+Q^{2}\right) \cos \left( \omega t\right) -2qQ\right] +\frac{i\hbar }{2}%
\ln \left( \frac{2\pi i\hbar \sin \left( \omega t\right) }{m\omega }\right)
\label{prop1}
\end{equation}%
does indeed solve Eq.(\ref{QHJ3}) with the appropriate boundary condition%
\footnote{%
In the limit $\omega \rightarrow 0$, insertion of Eq.(\ref{prop1}) into Eq.(%
\ref{QHJ4}) reproduces of course the kernel for the $1D$ free particle.}. It
is then immediate to see that the "well-ordered" solution of the operator
equation (\ref{QCT2})  will be:%
\begin{equation}
S=\frac{m\omega }{2\sin \left( \omega t\right) }\left[ \left( \widehat{q}%
^{2}+\widehat{Q}^{2}\right) \cos \left( \omega t\right) -2\widehat{q}%
\widehat{Q}\right] +\frac{i\hbar }{2}\ln \left( \frac{2\pi i\hbar \sin
\left( \omega t\right) }{m\omega }\right)   \label{prop2}
\end{equation}%
\begin{remark}
When substituting from Eq.(\ref{prop2}) into Eq.(\ref{QCT2}) we need to square the
derivative: $\partial S/\partial\widehat{q}=m\omega(\widehat{q}\cos(\omega
t)-\widehat{Q})/\sin(\omega t)$. This brings about terms that are not
"well-ordered". Bringing them in the correct order \cite{RS} requires using
the (exact) commutation relation: $\left[  \widehat{q},\widehat{Q}\right]
=-i\hbar\sin\left(  \omega t\right)  /m\omega$. The time derivative of the
last term in Eq.(\ref{prop2}) (see also footnote \ref{foot2}) does then the job of compensating for this operation.
\end{remark}

The short-time limit of Eq.(\ref{prop2}) is:%
\begin{equation}
S\underset{t\rightarrow 0}{\approx }\frac{m}{2t}\left( \widehat{q}^{2}+%
\widehat{Q}^{2}-2\widehat{q}\widehat{Q}\right) +\frac{i\hbar }{2}\ln \left(
\frac{2\pi i\hbar t}{m}\right)   \label{prop3}
\end{equation}%
which is the same as for the free particle.
The harmonic potential does not contribute to this short-time limit. As remarked in Ref.\cite{RS}, the same will happen more generally for any
nonsingular potential $V\left( \widehat{q}\right) $. So, the result (\ref%
{prop3}) will have a more general significance, going slightly beyond the case of
quadratic potentials. \

\subsection{Comments and Plan of the Paper.}\label{comments}
Going back now to our general discussion, the attempt to recover the original
$PDE$ from the equations of motion of the "rays", i.e. Hamilton's equations,
is what is usually called the "quantization problem".

This interplay between $PDE$'s, associated first-order $PDE$'s ($HJ$-type
equations) describing the propagation of "wave-fronts" and the corresponding,
"ray" $ODE$'s (Hamilton's equations) has been widely investigated in various
branches of theoretical Physics. Moreover, because of the
Hamiltonian-Lagrangian correspondence, also the calculus of variations appears
in this interplay.

>From all these remarks it should be clear that the $HJ$ theory is very rich in
analytic and geometric ideas, and that it unifies apparently diverse topics
like higher-order $PDE$'s, first-order $PDE$'s, $ODE$'s and the calculus of variations.

In this paper we will try and use some of our experience with Quantum
Mechanics to formulate and give a geometric\ presentation of many problems
which, born in a quantum setting, are of more general validity in the
framework of the Hamilton-Jacobi theory. Besides many original papers on this
vast subject, we shall rely on some work by A.Vinogradov \cite{VK}, some more recent work by Grabowski and Poncin \cite{GP}, a previous
paper of ours \cite{MMN} and recent paper on the Hamilton-Jacobi theory in a
Lagrangian setting \cite{CGMM}. We shall discuss the following topics:

-Reviewing briefly how the $HJ$ problem can be formulated in geometric terms
on the cotangent bundle $T^{\ast}Q$ of a smooth manifold $Q$, we shall discuss
how the same problem can be formulated on the tangent bundle $TQ$, hence in a
Lagrangian setting.

-How one can pose a generalized $HJ$ problem for differential operators of any
order, giving another coordinate-free characterization of the $HJ$ equation.

-We shall try and discuss to which extent, instead of differential
operators acting on functions, one can consider differential operators acting
on sections of vector bundles, thereby obtaining "wave-like" equations that
are not scalar, like the Pauli and the Dirac equations.

-Just as in Quantum Mechanics one poses a joint eigenvalue problem for two (or
more) observables, we shall discuss how one can pose a "joint $HJ$ problem"
for more than two functions on the cotangent bundle, and, finally,

-After "revisiting" briefly the geometrical formulation of the time-dependent
Hamilton-Jacobi theory, whose proper setting \cite{MMN} is on the cotangent bundle
$T^{\ast}\left(  Q\times\mathbb{R}\right)  $, we shall discuss how one can
obtain generalizations thereof when the action of the Abelian group
$\mathbb{R}$ is replaced with that of a general Lie group $\mathbb{G}$ (a
"$HJ$ problem on a Lie group", then).

\bigskip

\section{A Geometrical Setting for the Time-Independent Hamilton-Jacobi Theory
on the Cotangent and on the Tangent Bundles.}\label{TIHJ}
\subsection{Preliminaries.}
Let us begin by recalling \ \cite{MMN} some preliminary notions. The cotangent
bundle $T^{\ast}Q$ carries with it the canonical (exact) symplectic structure:%
\begin{equation}
\omega_{0}=d\theta_{0} \label{symp_form}%
\end{equation}
where, in local coordinates: $\theta_{0}=p_{i}dq^{i}$, and hence: $\omega
_{0}=dp_{i}\wedge dq^{i}$.

Let then $\alpha\in\mathfrak{X}^{\ast}\left(  Q\right)  $ be a one-form on the
base manifold $Q$. Again, in coordinates:%
\begin{equation}
\alpha\left(  q\right)  =\alpha_{i}\left(  q\right)  dq^{i} \label{one_form}%
\end{equation}

With the one-form $\alpha$ we can associate the map (the \textit{graph} of
$\alpha$):%
\begin{equation}
\varphi_{\alpha}:Q\rightarrow T^{\ast}Q\text{ \ }by:\text{ }Q\ni
q\mapsto\left(  q,\alpha\left(  q\right)  \right)  \in T^{\ast}Q \label{map1}%
\end{equation}
The \textit{image} of $\alpha$, $\Gamma\left[  \alpha\right]  $ will be
defined as: $\Gamma\left[  \alpha\right]  =\varphi_{\alpha}\left(  Q\right)
$. Of course: $\dim\Gamma\left[  \alpha\right]  =\dim Q=n$. It is clear from
Eq.(\ref{map1}) that:%
\begin{equation}
\pi\circ\varphi_{\alpha}=Id_{Q}%
\end{equation}
where: $\pi:T^{\ast}Q\rightarrow Q$ is the canonical projection.
This shows that \ $\varphi_{\alpha}$ gives a global section of
$T^{\ast}Q$. $\Gamma\left[  \alpha\right]  $ will be therefore an
$n$-dimensional transversal \cite{MMN} submanifold of $T^{\ast}Q$
and $\varphi_{\alpha}$ will be an embedding \cite{MSSV} of $Q$ into
$T^{\ast}Q$. Also, $\alpha$ can be recovered from the canonical
one-form $\theta_{0}$ via the pull-back:%
\begin{equation}
\varphi_{\alpha}^{\ast}\theta_{0}=\alpha\label{pull_back1}%
\end{equation}

We can consider also the map:%
\begin{equation}
\psi_{\alpha}=:\varphi_{\alpha}\circ\pi:T^{\ast}Q\rightarrow T^{\ast}Q\text{
\ }by:T^{\ast}Q\ni\left(  q,p\right)  \mapsto\left(  q,\alpha\left(  q\right)
\right)  \in T^{\ast}Q\text{ } \label{map2}%
\end{equation}
which is a base-invariant translation along the fibers that, for every $q\in
Q$, "shrinks" the whole fibre $T_{q}^{\ast}Q$ to the point $\alpha\left(
q\right)  $.\ Then, using Eq.(\ref{pull_back1}), one obtains at once:%
\begin{equation}
\psi_{\alpha}^{\ast}\omega_{0}=\pi^{\ast}d\alpha\label{pull-back2}%
\end{equation}
and we conclude \cite{MMN} that the graph of $\alpha$, besides being always
transversal to the fibers, will be also a Lagrangian submanifold
\cite{MMN,MSSV} of $T^{\ast}Q$ \textit{if and only if }$\alpha$\textit{ is
closed}. As discussed in Ref.\cite{MMN}, the converse is not true, i.e. a
transversal Lagrangian submanifold of $T^{\ast}Q$ which projects down to $Q$
under the canonical projection need not be the graph of a closed one-form.

Being closed, $\alpha$ will be locally exact, i.e.: $\alpha=dW$ for
some function $W\in\mathcal{F}\left(  Q\right)  $, at least locally.
Whether or not such a function exists globally will depend on
whether or not $H^{1}\left( Q\right)  $, the first de Rham
cohomology group of $Q$, is trivial.
\subsection{The $HJ$ Theory on the Cotangent Bundle. }\label{TIHJ2}
After these preliminaries, let us re-consider now the
time-independent $HJ$ equation, Eq. (\ref{HJ2}), for Hamilton's
characteristic function, i.e.:
\begin{equation}
H\left(  q;\frac{\partial W}{\partial q}\right)  =E \label{HJ5}%
\end{equation}

$W$ can be either a particular solution of the $HJ$ equation or a complete
integral (with all the possible intermediate cases in between). In the latter
case: $W=W\left(  q;a\right)  $ depending (as already discussed, in an
essential way) on $n$ additional variables collectively denoted as $a$.
Whenever necessary, we will denote as $W_{a}\left(  q\right)  $ the function
that obtains by keeping the $a$'s constant. Hence: $W_{a}\in\mathcal{F}\left(
Q\right)  $ $\forall a$.

It is clear that, considering the graph of the exact one-form $dW_{a}$, the
image of $dW_{a}$ is a Lagrangian submanifold and Eq.(\ref{HJ5}) can be
rewritten as \cite{MMN}:%
\begin{equation}
dW_{a}^{\ast}\left(  H-E\right)  =0 \label{HJ6}%
\end{equation}
or, equivalently, as:%
\begin{equation}
dW_{a}^{\ast}\left(  dH\right)  =0 \label{HJ7}%
\end{equation}

As it stands, Eq.(\ref{HJ6}) looks just like a different way of rewriting the
standard $HJ$ equation in a different language. However it allows for a deeper
geometrical interpretation of the $HJ$ theory, and allows also for some
interesting generalizations.

First of all, the image $\Gamma_{a}$ of $dW_{a}$ is a regular submanifold
\cite{MSSV} of $T^{\ast}Q$. As such, it can be described, locally at least, as
the zero-level set of $n$ independent functions $f_{1a},...,f_{na}%
\in\mathcal{F}\left(  T^{\ast}Q\right)  $, i.e. such that:%
\begin{equation}%
\begin{array}
[c]{c}%
df_{1a}\wedge...\wedge df_{na}\neq0\\
\Gamma_{a}=\left\{  m=\left(  q,p\right)  \in T^{\ast}Q|\text{
}f_{ja}\left( m\right), j=0,\text{ }1,...,n\right\}
\end{array}
\label{level}%
\end{equation}
and:%
\begin{equation}
\left\{  f_{ia},f_{ja}\right\}  \left(  m\right)  =0 \label{PB1}%
\end{equation}
where $\{.,.\}$ is the Poisson bracket in the space of functions on
$T^{\ast}Q$ associated with the symplectic form $\omega_{0}$.\\
Eq.(\ref{PB1}) implies that the Hamiltonian vector fields $X_{j}$
associated
with the $f_{j}$'s, i.e.:%
\begin{equation}
i_{X_{ja}}\omega_{0}=-df_{ja},\text{ }j=1,...,n \label{Hamvec1}%
\end{equation}
are all tangent to the submanifold (indeed \cite{MMN} they span the tangent
space $T_{m}\Gamma_{a}$ at each $m\in\Gamma_{a}$). Moreover, denoting by
$X_{H}$ the Hamiltonian vector field that describes the dynamics, we obtain,
contracting both sides of Eq.(\ref{HJ7}) with the $X_{ja}$'s:%
\begin{equation}%
\begin{array}
[c]{c}%
0=\left[  i_{X_{ja}}\left(  dW_{a}^{\ast}\left(  dH\right)  \right)  \right]
\left(  m\right)  =\left[  i_{X_{ja}}dH\right]  \left(  m\right)  =\\
=\left(  L_{X_{ja}}H\right)  \left(  m\right)  =-\left(  L_{X_{H}}%
f_{ja}\right)  \left(  m\right)  =\left\{  H,f_{ja}\right\}  \left(  m\right)
\end{array}
\label{tangent}%
\end{equation}
which proves that $X_{H}$ is tangent to $\Gamma_{a}\forall a$. Of course
$X_{H}$ is also tangent to the $\left(  2n-1\right)  $-dimensional energy
surface\footnote{Remember that, in the time-independent case, the energy $E$
must be included among the parameters on which a complete integral depends,
leaving actually only $n-1$ independent parameters.}:%
\begin{equation}
\Sigma_{E}=\left(  H-E\right)  ^{-1}\left(  0\right)  \label{en_surf}%
\end{equation}
If $dH\neq0$\footnote{Critical points of $H$, being invariant sets for the
dynamics, can be handled separately.}, $\Sigma_{E}$ is a regular submanifold
that we can assume without loss of generality to be also connected. If it is
not, we can always restrict the discussion to each connected component
separately. Each $\Gamma_{a}$ is contained in the energy surface
(\ref{en_surf}) and, in general \cite{MMN}, they will provide a $\left(
n-1\right)  $-parameter foliation of the energy surface, with the dynamical
vector field $X_{H}$ being tangent to all the leaves of the foliation. This is
basically the geometrical content of the fact that, besides the energy $E$, a
complete integral depends in an essential way on $n-1$ additional parameters.

This formulation of the $HJ$ problem suffers however of some limitations, as
the following example shows.

\textbf{Example 1. }\textit{Consider: $Q=S^{1}$ with coordinate $q,0\leq q<2\pi$. The cotangent bundle can
be given coordinates $\left(  q,p\right)  $, with $p\in\mathbb{R}$, and can be
viewed as a cylinder. In this case: $\theta_{0}=pdq$ and: $\omega_{0}=dp\wedge
dq$ are both well-defined. The vector field: $X=p\partial/\partial q$ is
Hamiltonian with the Hamiltonian: $H=p^{2}/2$, and the "energy surfaces" are
pairs of circles on the cylinder: $\Sigma_{E}=\left(  q,\pm\sqrt{2E}\right)
$, $E\geq0$. The associated $HJ$ equation: $\left(  \partial S/\partial
q\right)  ^{2}=E$ \ has of course no global solutions for $E>0$, but the
energy surfaces are nonetheless the graphs of the closed but not exact
one-forms: $\alpha_{E,\pm}=\pm\sqrt{2E}dq$, and the equation: $\varphi
_{\alpha}^{\ast}\left(  H-E\right)  =0$ for an unknown closed one-form
$\alpha$ is globally defined and has precisely $\alpha_{E,\pm}$ as solutions.}

This suggests that one relaxes the requirement \ that the pull-back of $H-E$
in Eq.(\ref{HJ6}) be via an exact one-form, replacing it with the weaker
request that it be via a closed but not necessarily exact one-form, i.e. that
one replaces Eq.(\ref{HJ6}) with:%
\begin{equation}
\alpha^{\ast}\left(  H-E\right)  =0,\text{ }\alpha\in\mathfrak{X}^{\ast
}\left(  Q\right)  ,\text{ }d\alpha=0 \label{HJ8}%
\end{equation}
The graph of any such form will be of course again a transversal Lagrangian
submanifold, and we can re-formulate the "geometric $HJ$ problem" as follows:

\begin{itemize}
\item A \textit{solution} of the $HJ$ equation for a given Hamiltonian and a
given energy is a transversal Lagrangian submanifold within the energy
surface, obtained as the graph of a closed one-form on $Q$, and

\item A \textit{complete integral} is a foliation of $T^{\ast}Q$ by such
solutions for all physically accessible values of the energy, which implies
also a foliation of each energy surface as well.
\end{itemize}

Note that (cfr. Eq.(\ref{pull_back1})), as: $\alpha^{\ast}\omega_{0}%
=\alpha^{\ast}d\theta_{0}=d(\alpha^{\ast}\theta_{0})=d\alpha$, Eq.(\ref{HJ8})
can be replaced by the equivalent one:%
\begin{equation}
\alpha^{\ast}\left(  H-E\right)  =0,\text{ }\alpha\in\mathfrak{X}^{\ast
}\left(  Q\right)  ,\text{ }\alpha^{\ast}\omega_{0}=0 \label{HJ9}%
\end{equation}

If, as a further step, one gives up the requirement of transversality, one can
pose a "$HJ$ problem" (no more a "$HJ$ equation") consisting in the search for
foliations of each energy surface simply by Lagrangian submanifolds (that,
being not necessarily transversal, can exhibit caustics \cite{AVG}). This has
the advantage that the full set of canonical symmetries can be implemented as
symmetries of the $HJ$ problem (i.e. maps that map solutions into solutions).
We will not insist on this point, but refer rather to the literature
\cite{MMN} for a more complete discussion.
\subsection{The $HJ$ Theory on the Tangent Bundle. }
We turn now briefly to the Lagrangian context. The relevant carrier space is
now the tangent bundle $TQ$, with local coordinates $\left(  q^{i}%
,u^{i}\right)  ,i=1,...,n$, which carries no pre-assigned symplectic structure
but, if a Lagrangian $\mathcal{L}\in\mathcal{F}\left(  TQ\right)  $ (assumed
here to be regular\footnote{We refer to Ref. \cite{MSSV} for a discussion of
the case of singular Lagrangians.}) is given, can be endowed with a
\textit{Lagrangian} symplectic structure $\omega_{\mathcal{L}}$ defined by:
\cite{AM,Ar,MSSV}:%
\begin{equation}
\omega_{\mathcal{L}}=d\theta_{\mathcal{L}};\text{ }\theta_{\mathcal{L}}%
=\frac{\partial\mathcal{L}}{\partial u^{i}}dq^{i}%
\end{equation}
again, for simplicity, in local coordinates. The Euler-Lagrange equations can
be put in "Hamiltonian" form as:%
\begin{equation}
i_{\Gamma_{\mathcal{L}}}\omega_{\mathcal{L}}=-dE_{\mathcal{L}}%
\end{equation}
where $\Gamma_{\mathcal{L}}$ is the second-order \cite{MSSV} vector field
describing the dynamics, the "energy function" $E_{\mathcal{L}}$ is given by:%
\begin{equation}
E_{\mathcal{L}}=\left(  L_{\Delta}-1\right)  \mathcal{L}%
\end{equation}
and, finally, $\Delta$ is the dilation (Liouville) field along the fibers:%
\begin{equation}
\Delta=u^{i}\frac{\partial}{\partial u^{i}}%
\end{equation}

Sections (actually, global sections) of the tangent bundle are now provided by
vector fields on the base manifold, just as one-forms did the same job for the
cotangent bundle. If $X\in\mathfrak{X}\left(  Q\right)  $ is any such vector
field, given in local coordinates by: $X=X^{i}\left(  q\right)  \partial
/\partial q^{i}$, then $X$ will define the map\footnote{With some abuse of
notation, we are using here too the same symbol to denote the vector field and
the associated map.}:%
\begin{equation}
X:Q\rightarrow TQ\text{ }by:\text{ }q^{i}\mapsto\left(  q^{i},X^{i}\left(
q\right)  \right)
\end{equation}
satisfying:%
\begin{equation}
\pi\circ X=Id_{Q}%
\end{equation}
where: $\pi:TQ\rightarrow Q$ is the canonical projection, and hence $X$ is a section of
$TQ$.

With reference to Eq.(\ref{HJ9}) we can then define a \textit{Lagrangian }$HJ$
\textit{problem} as the search for all vector fields $X\in\mathfrak{X}\left(
Q\right)  $ such that:%
\begin{equation}
X^{\ast}\left(  E_{\mathcal{L}}-E\right)  =0\text{ \ }and:\text{ \ }X^{\ast
}\omega_{\mathcal{L}}=0\label{HJ10}%
\end{equation}

As such, the $HJ$ problem on the tangent bundle can be viewed as the search,
instead of a single function as in the case of $T^{\ast}Q,$ of a
"vector-valued function" on $Q$.

As a (Lagrangian) counterpart of the Remark that was made in Sect.\ref{Intro},
we have now the following \cite{CGMM}
\begin{remark}
If: $X\in\mathfrak{X}\left(  Q\right)  ,$ $X:Q\rightarrow TQ$ is a solution of
Eq.(\ref{HJ10}) and $\gamma:\mathbb{R}\rightarrow Q$ is an integral curve of
$X$, i.e.: $\overset{\mathbf{\cdot}}{\gamma}=X\circ\gamma$, then:
$\overset{\mathbf{\cdot}}{\gamma}:\mathbb{R}\rightarrow TQ$ solves the
Euler-Lagrange equations for the Lagrangian $\mathcal{L}$, i.e.:
\end{remark}%
\begin{equation}
X\circ\gamma=\overset{\mathbf{\cdot}}{\gamma}\Longrightarrow\Gamma
_{\mathcal{L}}\circ\overset{\mathbf{\cdot}}{\gamma}=\overset{\mathbf{\cdot}%
}{\overline{X\circ\gamma}}\label{HJ11}%
\end{equation}

The converse statement, however (i.e. if $\ $the integral curves $\gamma$ of a
vector field $X\in\mathfrak{X}\left(  Q\right)  $ are such that $\overset
{\mathbf{\cdot}}{\gamma}\ $ are integral curves for the Lagrangian vector
field $\Gamma_{\mathcal{L}}$ for a given Lagrangian $\mathcal{L}$, then $X$ is
a solution of Eq.(\ref{HJ10})) need not be true, as the following example
\cite{CGMM} shows.

\textbf{Example 2. }\textit{The dynamics of the free particle in }%
$\mathbb{R}^{2}\mathit{\ }$\textit{can be described by the regular
Lagrangian:}%
\begin{equation}
\mathcal{L=}\frac{1}{2}\left[  \left(  u^{1}\right)  ^{2}+\left(
u^{2}\right)  ^{2}\right]
\end{equation}
\textit{with associated geometrical objects:}%
\begin{equation}
\Gamma_{\mathcal{L}}=u^{1}\frac{\partial}{\partial q^{1}}+u^{2}\frac{\partial}{\partial q^{2}};  E_{\mathcal{L}}=\frac{1}{2}\left[  \left(
u^{1}\right)  ^{2}+\left(  u^{2}\right)  ^{2}\right]  \text{\ }%
\end{equation}
\textit{and:}%
\begin{equation}
\theta_{\mathcal{L}}=u^{1}dq^{1}+u^{2}dq^{2};\text{ \ }\omega_{\mathcal{L}%
}=du^{1}\wedge dq^{1}+du^{2}\wedge dq^{2}%
\end{equation}
\textit{The two-parameter family of vector fields:}%
\begin{equation}
X=k\frac{\partial}{\partial q^{1}}+\frac{kq^{2}-l}{q^{1}}\frac{\partial
}{\partial q^{2}}%
\end{equation}
\textit{satisfies the assumptions of the above (putative) converse statement, but:}%
\begin{equation}
X^{\ast}\left(  \omega_{\mathcal{L}}\right)  =-\frac{kq^{2}-l}{\left(
q^{1}\right)  ^{2}}dq^{1}\wedge dq^{2}\neq0
\end{equation}
\textit{and:}%
\begin{equation}
X^{\ast}\left(  E_{\mathcal{L}}\right)  =\frac{1}{2}\left[  k^{2}+\left(
\frac{kq^{2}-l}{q^{1}}\right)  ^{2}\right]  \Rightarrow d\left(  X^{\ast
}\left(  E_{\mathcal{L}}\right)  \right)  \neq0
\end{equation}
\textit{Hence, both conditions of Eq.(\ref{HJ10}) will be violated.}

Guided by this example we will stick to the definition of the
\textit{Lagrangian Hamilton-Jacobi problem }as defined by Eqs.(\ref{HJ10}),
i.e. as the search for all vector fields $X\in\mathfrak{X}\left(  Q\right)  $
such that (using again the same notation for the associated maps $Q\rightarrow
TQ$):%
\begin{equation}
X^{\ast}\omega_{\mathcal{L}}=d(X^{\ast}E_{\mathcal{L}})=0 \label{HJ12}%
\end{equation}
The first of these equations implies, of course, that $\operatorname{Im}X$
\ be a Lagrangian submanifold of $TQ$. Furthermore, as: $0=$ $X^{\ast}%
\omega_{\mathcal{L}}=X^{\ast}(d\theta_{\mathcal{L}})=d\left(  X^{\ast}%
\theta_{\mathcal{L}}\right)  $, every point has an open neighborhood $U\subset
Q$ where there is a function $W\in\mathcal{F}\left(  U\right)  $ such that:
$X^{\ast}\theta_{\mathcal{L}}=dW$ in $U$.

We have already shown in a previous Remark that if $X$ is a solution of the
Lagrangian $HJ$ problem, then it satisfies Eq.(\ref{HJ11}) (while the converse
is not true)\footnote{In Ref.\cite{CGMM} the problem of finding the solutions
of Eq.(\ref{HJ11}) is also called the \textit{generalized} Lagrangian $HJ$
problem.}. It has been proved in Ref.\cite{CGMM} that $X$ is a solution of
Eq.(\ref{HJ10}) iff the following diagram:%
\begin{equation}%
\begin{array}
[c]{ccccc}
&  &  &  & \\
& TQ & \overset{\Gamma_{\mathcal{L}}}{\longrightarrow} & T\left(  TQ\right)
& \\
X & \uparrow &  & \uparrow & TX\\
& Q & \underset{X}{\longrightarrow} & TQ & \\
&  &  &  &
\end{array}
\end{equation}
commutes, i.e. iff:%
\begin{equation}
\Gamma_{\mathcal{L}}\circ X=TX\circ X\label{related}%
\end{equation}
Eq.(\ref{related}) is a $PDE$ for the unknown vector field, or "vector-valued
function", $X$ which replaces the $PDE$ for the scalar function $W$. Once a
solution is found, one has however to check whether or not it satisfies the
conditions (\ref{HJ12}) as well, in order for it to be a genuine solution of
the Lagrangian $HJ$ problem.

It may be useful to derive the expression for the $PDE$ (\ref{related}) in
local coordinates. We can write $X$ and $\Gamma_{\mathcal{L}}$ as:%
\begin{equation}%
\begin{array}
[c]{c}%
X=X^{i}\frac{\partial}{\partial q^{i}};\text{ }X^{i}=X^{i}\left(  q\right)  \\
\Gamma_{\mathcal{L}}=u^{i}\frac{\partial}{\partial q^{i}}+a^{i}\frac{\partial
}{\partial u^{i}}%
\end{array}
\end{equation}
where:%
\begin{equation}
a^{i}=a^{i}\left(  q,u\right)  =H^{ij}\left\{  \frac{\partial\mathcal{L}%
}{\partial q^{j}}-\frac{\partial^{2}\mathcal{L}}{\partial u^{j}\partial q^{k}%
}u^{k}\right\}
\end{equation}
and $H^{ij}$ is the inverse of the Hessian matrix:%
\begin{equation}
H_{ij}=\frac{\partial^{2}\mathcal{L}}{\partial u^{i}\partial u^{j}}%
\end{equation}
A direct calculation shows that:%
\begin{equation}
\left(  TX\circ X-\Gamma_{\mathcal{L}}\circ X\right)  \left(  q\right)
=\left(  \frac{\partial X^{i}}{\partial q^{j}}X^{j}-a^{i}\left(  q,X\right)
\right)  \frac{\partial}{\partial u^{i}}%
\end{equation}
This is a vertical vector field along $X$ whose vanishing implies:%
\begin{equation}
\frac{\partial X^{i}}{\partial q^{j}}X^{j}-a^{i}\left(  q,X\right)  =0,\text{
}i=1,...,n\label{PDE2}%
\end{equation}
and this is the required local form of the $PDE$ (\ref{related}).

\bigskip

One of the main results of Ref.\cite{CGMM} can now be rephrased as follows:

\textbf{Proposition.} \textit{The following statements are equivalent:}

\begin{enumerate}
\item $X$ \textit{is a solution of the Lagrangian }$HJ$ \textit{problem.}

\item \textit{Besides being a Lagrangian submanifold of }$TQ$\textit{,
}$\operatorname{Im}X$ \textit{is also invariant under the dynamics represented
by }$\Gamma_{\mathcal{L}}$, \textit{i.e.} $\Gamma_{\mathcal{L}}$ \textit{is
everywhere tangent to }$\operatorname{Im}X$.

\item \textit{The integral curves of }$\Gamma_{\mathcal{L}}$ \textit{with
initial conditions on }$\operatorname{Im}X$ \textit{project onto the integral
curves of }$X$.
\end{enumerate}

In the context of the tangent bundle, the notion of a complete solution of the
Lagrangian $HJ$ problem can be posed as follows:

\textbf{Definition.} \textit{A \underline{\textit{complete solution}} of the
Lagrangian Hamilton-Jacobi problem is provided by a family of solutions
}$\left\{  X_{\lambda}\right\}  _{\lambda\in\Lambda}$\textit{, }$\Lambda$
\textit{an open set in }$\mathbb{R}^{n}$\textit{, such that the map:}%
\begin{equation}
\Phi:Q\times\Lambda\rightarrow TQ\text{ \ }by:\text{ }\Phi\left(
q,\lambda\right)  =X_{\lambda}\left(  q\right)
\end{equation}
\textit{is a local diffeomorphism.}

It follows from the definition that a complete solution yields a
foliation\footnote{This is ensured by the request that $\Phi\left(
q,\lambda\right)  $ be a (local) diffeomorphism.} of $TQ$ with leaves
transversal to the fibers, and that the Lagrangian vector field $\Gamma
_{\mathcal{L}}$ is tangent to the fibers of the foliation.

\textbf{Example 3.} \textit{Consider the two-dimensional harmonic oscillator
\ with the standard Lagrangian:}%
\begin{equation}
\mathcal{L}=\frac{1}{2}\left[  \left(  u^{1}\right)  ^{2}+\left(
u^{2}\right)  ^{2}-\left(  q^{1}\right)  ^{2}-\left(  q^{2}\right)
^{2}\right]\label{Lag1}
\end{equation}

\textit{The dynamical vector field is:}%
\begin{equation}
\Gamma_{\mathcal{L}}=u^{1}\frac{\partial}{\partial q^{1}}+u^{2}\frac{\partial
}{\partial q^{2}}-q^{1}\frac{\partial}{\partial u^{1}}-q^{2}\frac{\partial
}{\partial u^{2}}%
\end{equation}
\textit{and:}%
\begin{equation}%
\begin{array}
[c]{c}%
E_{\mathcal{L}}=\frac{1}{2}\left[  \left(  u^{1}\right)  ^{2}+\left(
u^{2}\right)  ^{2}-\left(  q^{1}\right)  ^{2}-\left(  q^{2}\right)
^{2}\right]  \\
\omega_{\mathcal{L}}=du^{1}\wedge dq^{1}+du^{2}\wedge dq^{2}%
\end{array}
\end{equation}

\textit{It is known that the functions:}%
\begin{equation}%
\begin{array}
[c]{c}%
f_{0}=u^{1}u^{2}+q^{1}q^{2}\\
f_{1}=\left(  u^{1}\right)  ^{2}+\left(  q^{1}\right)  ^{2}\\
f_{2}=\left(  u^{2}\right)  ^{2}+\left(  q^{2}\right)  ^{2}\\
f_{3}=q^{1}u^{2}-q^{2}u^{1}%
\end{array}
\end{equation}
\textit{are all constants of the motion, not functionally independent, of
course. Let, e.g., }$f_{0}=C,f_{1}=2E_{1},f_{2}=2E_{2},f_{3}=l$\textit{. In
particular, }$f_{1}$\textit{ and }$f_{2}$\textit{ are in involution:
}$\left\{  f_{1},f_{2}\right\}  =0.$\textit{ The }$PDE$\textit{ equations
(\ref{PDE2}) read now:}%
\begin{equation}
\frac{\partial X^{i}}{\partial q^{j}}X^{j}=-q^{i},\text{ }i=1,2\label{PDE3}
\end{equation}
\textit{and it is easy to check that the four two-parameter families
of vector fields\footnote{It is easy to recognize that the
components of the vector fields are obtained by expressing $u^{1,2}$
in terms of the coordinates on the
base manifold and of the two parameters $E_{1,2}$.}:\ }%
\begin{equation}
X_{E_{1},E_{2}}=\pm\sqrt{2E_{1}-\left(  q^{1}\right)
^{2}}\frac{\partial }{\partial q^{1}}\pm\sqrt{2E_{2}-\left(
q^{2}\right)  ^{2}}\frac{\partial
}{\partial q^{2}}\label{family}%
\end{equation}
\textit{do indeed solve them. As:}%
\begin{equation}
\left(  X_{E_{1}E_{2}}\right)  ^{\ast}\omega_{\mathcal{L}}=\pm d\sqrt
{2E_{1}-\left(  q^{1}\right)  ^{2}}\wedge dq^{1}\pm d\sqrt{2E_{2}-\left(
q^{2}\right)  ^{2}}\wedge dq^{2}=0
\end{equation}
\textit{and:}%
\begin{equation}%
\begin{array}
[c]{c}%
\left(  X_{E_{1},E_{2}}\right)  ^{\ast}dE_{\mathcal{L}}=\\
=\sqrt{2E_{1}-\left(  q^{1}\right)  ^{2}}d\sqrt{2E_{1}-\left(
q^{1}\right) ^{2}}+\sqrt{2E_{2}-\left(  q^{2}\right)
^{2}}d\sqrt{2E_{2}-\left( q^{2}\right)
^{2}}+q^{1}dq^{1}+q^{2}dq^{2}=0
\end{array}
\end{equation}
\textit{each family (\ref{family}) is indeed a complete solution of
the Lagrangian }$HJ$ \textit{problem for the two-dimensional
harmonic oscillator with the standard Lagrangian (\ref{Lag1}) .}
\begin{remark}
It is useful to stress here that the $PDE$ (\ref{PDE2})
(or, for that matter, Eq.(\ref{PDE3}) in the case of the harmonic oscillator),
which defines \cite{CGMM}  the "generalized" Lagrangian $HJ$ problem, depends
on the Lagrangian only through the vector field $\Gamma_{\mathcal{L}%
}$, while the $HJ$ problem that we are discussing here depends also on
additional structures derived from the Lagrangian, namely the symplectic form
$\omega_{\mathcal{L}}$ and the "energy function" $E_{\mathcal{L}}$.
\end{remark}
It is known \cite{MFLMR} that the so-called "Inverse Problem in the Calculus
of Variations", i.e. the problem of whether or not a dynamics described by a
given second-order vector field on $TQ$ admits of a Lagrangian description can
have no solutions at all, only one solution\footnote{Apart from the addition
\cite{MFLMR} of essentially trivial "gauge terms".}, or more than one
solution, leading then to (genuinely) alternative Lagrangian descriptions for
a given dynamics.

If this is the case, vector fields that are solutions of the Lagrangian $HJ$
problem (or families thereof, yielding complete solutions) for a
\textit{given} Lagrangian need not be such for an alternative Lagrangian
description, while remaining, according to what has just been said, solutions
(or complete solutions) of the \textit{generalized }\ $HJ$ problem, as the
following example shows.

\textbf{Example 4. }\textit{Consider again the two-dimensional harmonic
oscillator, but now with the alternative Lagrangian \cite{MFLMR}:}%
\begin{equation}
\mathcal{L}_{1}=\frac{1}{2}\left[  \left(  u^{1}\right)  ^{2}-\left(
u^{2}\right)  ^{2}-\left(  q^{1}\right)  ^{2}+\left(  q^{2}\right)
^{2}\right]  \label{Lag2}%
\end{equation}
\textit{The associated structures will be now:}%
\begin{equation}%
\begin{array}
[c]{c}%
E_{\mathcal{L}_{1}}=\frac{1}{2}\left[  \left(  u^{1}\right)  ^{2}-\left(
u^{2}\right)  ^{2}+\left(  q^{1}\right)  ^{2}-\left(  q^{2}\right)
^{2}\right]  \\
\omega_{\mathcal{L}_{1}}=du^{1}\wedge dq^{1}-du^{2}\wedge dq^{2}%
\end{array}
\end{equation}
\textit{while, of course, the dynamical vector field and the }$PDE$
\textit{(\ref{PDE3}) will be the same.}

\textit{The vector fields (\ref{family}) will satisfy again }$\left(
X_{E_{1}E_{2}}\right)  ^{\ast}\omega_{\mathcal{L}_{1}}=0$ \textit{as well as:
}$\left(  X_{E_{1},E_{2}}\right)  ^{\ast}dE_{\mathcal{L}_{1}}=0$ \textit{and
hence the }$X_{E_{1}E_{2}}$\textit{'s will provide a complete solution of the
}$HJ$ \textit{problem for the Lagrangian (\ref{Lag2}) as well.}

I\textit{f one considers instead the Lagrangian \cite{MFLMR}:}%
\begin{equation}
\mathcal{L}_{2}=u^{1}u^{2}-q^{1}q^{2}\label{Lag3}%
\end{equation}
\textit{we have:}%
\begin{equation}%
\begin{array}
[c]{c}%
E_{\mathcal{L}_{2}}=u^{1}u^{2}+q^{1}q^{2}\\
\omega_{\mathcal{L}_{2}}=du^{2}\wedge dq^{1}+du^{1}\wedge dq^{2}%
\end{array}
\end{equation}
\textit{and one finds, e.g.:}%
\begin{equation}
\left(  X_{E_{1}E_{2}}\right)  ^{\ast}\omega_{\mathcal{L}_{2}}=\left[
\pm\frac{q^{2}}{\sqrt{2E_{2}-\left(  q^{2}\right)  ^{2}}}\pm\frac{q^{1}}%
{\sqrt{2E_{1}-\left(  q^{1}\right)  ^{2}}}\right]  dq^{1}\wedge dq^{2}\neq0
\end{equation}
(\textit{as well as: }$\left(  X_{E_{1}E_{2}}\right)  ^{\ast}\left(
dE_{\mathcal{L}_{2}}\right)  \neq0$\textit{). Hence, the }$X_{E_{1}E_{2}}%
$\textit{'s will be no more solutions of the }$HJ$ \textit{problem for \ the
alternative Lagrangian (\ref{Lag3}).}

\section{The Generalized Hamilton-Jacobi Problem for Differential Operators.}

\label{DiffOp}

\subsection{Differential Operators and Principal Symbols.\label{DOPS}}

In order to deal with differential operators on manifolds, we will first
review differential operators on $\mathbb{R}^{n}$. We will do this by
providing an algebraic characterization that will allow us to deal with
differential operators on arbitrary manifolds.

We consider then the algebra $\mathcal{A}=\mathcal{F}\left(  \mathbb{R}%
^{n}\right)  $ of smooth functions on $\mathbb{R}^{n}$. A \textit{differential
operator of degree at most }$k$ is defined as a linear map: $D^{\left(
k\right)  }:\mathcal{A\rightarrow A}$ of the form:%
\begin{equation}
D^{\left(  k\right)  }=%
{\displaystyle\sum\limits_{\left\vert \sigma\right\vert \leq k}}
g_{\sigma}\frac{\partial^{\left\vert \sigma\right\vert }}{\partial x_{\sigma}%
};\text{ }g_{\sigma}\in\mathcal{A} \label{diffop1}%
\end{equation}
where we have introduced multi-indices: $\sigma=\left(  i_{1},i_{2}%
,...,i_{n}\right)  ,$ $\left\vert \sigma\right\vert =i_{1}+i_{2}+...+i_{n}$,
and:%
\begin{equation}
\frac{\partial^{\left\vert \sigma\right\vert }}{\partial x_{\sigma}}%
=\frac{\partial^{\left\vert \sigma\right\vert }}{\partial x_{1}^{i_{1}%
}\partial x_{2}^{i_{2}}...\partial x_{n}^{i_{n}}}%
\end{equation}

It is possible to give an algebraic characterization, appropriate to arbitrary
manifolds, in the following way. With functions $f\in\mathcal{A}$ we associate
differential operators $\widehat{f}$ of order zero that act by multiplication,
i.e.: $\widehat{f}g=:fg$ on all smooth functions. We notice that the
commutator bracket gives:%
\begin{equation}
\left[  \frac{\partial}{\partial x_{i}},\widehat{f}\right]  =\widehat
{\frac{\partial f}{\partial x_{i}}}%
\end{equation}
and, more generally:%
\begin{equation}
\left[  \frac{\partial^{\left\vert \sigma\right\vert }}{\partial x_{\sigma}%
},\widehat{f}\right]  =%
{\displaystyle\sum\limits_{\tau+\nu=\sigma}}
c_{\tau}\frac{\partial^{\left\vert \tau\right\vert }f}{\partial x_{\tau}}%
\frac{\partial^{\left\vert \nu\right\vert }}{\partial x_{\nu}}%
\end{equation}
for $\left\vert \tau\right\vert >0$ (strictly) and some set of constants
$c_{\tau}$. It follows then easily that:%
\begin{equation}
\left[  D^{\left(  k\right)  },\widehat{f}\right]  =%
{\displaystyle\sum\limits_{\left\vert \sigma\right\vert \leq k}}
g_{\sigma}\left[  \frac{\partial^{\left\vert \sigma\right\vert }}{\partial
x_{\sigma}},\widehat{f}\right]
\end{equation}
is a differential operator of degree at most $k-1$. Iterating the procedure
for a set of $k+1$ functions $f_{0},f_{1},...,f_{k}$, we find:%
\begin{equation}
\left[  ...\left[  \left[  D^{\left(  k\right)  },\widehat{f}_{0}\right]
,\widehat{f}_{1}\right]  ,...,\widehat{f}_{k}\right]  =0 \label{diffop2}%
\end{equation}

The converse statement holds also true, namely, a linear operator
which  does not increase the support and satisfying the property (\ref{diffop2}) on any set of $k+1$ elements
in $\mathcal{A}$ will be a differential operator \cite{As}. By means
of this algebraic characterization it is then possible to define
differential operators as linear maps satisfying the property
(\ref{diffop2}).

We notice that:

\begin{itemize}
\item
\begin{equation}
\left[  D^{\left(  k\right)  },D^{\left(  j\right)  }\right]  =D^{\left(
k+j-1\right)  }\text{ \ },\text{ }k+j\geq1
\end{equation}

\item Setting: $k=j=1$ we obtain that differential operators of degree at most
one are a subalgebra and that $\mathcal{A}$, as an Abelian subalgebra of
operators of degree zero, is an invariant subalgebra thereof.

\item Differential operators of degree at most one are derivations of
$\mathcal{A}$ if they are zero on constants.

\item Derivations are a subalgebra (the subalgebra of \textit{homogeneous}
differential operators of degree one).
\end{itemize}

\begin{remark} \textit{Considering the algebra }$A$\textit{ of smooth functions and the
algebra }$DerA$\textit{ of the derivations on }$A$\textit{, and noticing that:
}$\left[  X_{1}+f_{1},X_{2}+f_{2}\right]  =\left[  X_{1},X_{2}\right]
+L_{X_{1}}f_{2}-L_{X_{2}}f_{1}$\textit{, }$X_{1},X_{2}\in DerA,$\textit{
}$f_{1},f_{2}\in A$\textit{, we can form a semi-direct product of Lie algebras
by setting:}%
\begin{equation}
\left[  \left(  X_{1},f_{1}\right)  ,\left(  X_{2},f_{2}\right)  \right]
=\left(  \left[  X_{1},X_{2}\right]  ,L_{X_{1}}f_{2}-L_{X_{2}}f_{1}\right)
\end{equation}
\textit{The enveloping algebra of this Lie algebra will be isomorphic with the
algebra of differential operators.}
\end{remark}
If $D$ is a differential operator, then for any two functions $f$ and $g$:%
\begin{equation}
\left[  \widehat{f},\left[  \widehat{g},D\right]  \right]  =\left[
\widehat{g},\left[  \widehat{f},D\right]  \right]  \label{symm}%
\end{equation}

It follows then that:

\textbf{Proposition. }\textit{If }$D^{\left(  k\right)  }$ \textit{is a
differential operator of degree at most }$k$\textit{, the expression:}%
\begin{equation}
\left[  ...\left[  D^{\left(  k\right)  },f_{1}\right]  ...f_{k}\right]
\end{equation}
\textit{is a function which is symmetric with respect to all the permutations
 of }$f_{1},...,f_{k}$.$\blacksquare$

We can set up an equivalence relation $"\simeq"$ among differential operators
 of the same degree, say $k$, by saying that $D_{1}^{\left(  k\right)  }\simeq
D_{2}^{\left(  k\right)  }$ iff:%
\begin{equation}
\left[  ...\left[  D_{1}^{\left(  k\right)  },f_{1}\right]  ...f_{k}\right]
=\left[  ...\left[  D_{2}^{\left(  k\right)  },f_{1}\right]  ...f_{k}\right]
\text{ \ \ }\forall f_{1},...,f_{k} \label{PS1}%
\end{equation}
The equivalence class is what is called the \textit{principal symbol} of the
differential operator. The set of principal symbols is a commutative algebra,
this following from the fact that two operators of degree $k$ are in the same
equivalence class iff their difference is a differential operator of degree
$<k$. The set of the symbols of the differential operators of degree $k$ will
be denoted as $\mathcal{S}^{\left(  k\right)  }\left(  Q\right)  $. The action
on a set \ $\mathbf{f}=\left(  f_{1},..,f_{k}\right)  $ of function of the
principal symbol of an operator $D^{\left(  k\right)  }$ of order $k$ will be
denoted as $\sigma^{P}\left(  D^{\left(  k\right)  }\right)  \left(
\mathbf{f}\right)  $ and, of course:%
\begin{equation}
\sigma^{P}\left(  D^{\left(  k\right)  }\right)  \left(  \mathbf{f}\right)
=\left[  ...\left[  D^{\left(  k\right)  },f_{1}\right]  ...f_{k}\right]
\label{PS2}%
\end{equation}
We note that:%
\begin{equation}
\left(  \sigma^{P}\left(  D^{\left(  k\right)  }\right)  \sigma^{P}\left(
D^{\left(  m\right)  }\right)  \right)  \left(  \mathbf{f}\right)  =\sigma
^{P}\left(  D^{\left(  k\right)  }\cdot D^{\left(  m\right)  }\right)  \left(
\mathbf{f}\right)
\end{equation}
with, now: $\mathbf{f}=\left(  f_{1},...,f_{k+m}\right)  $, which shows also that: $\sigma^{P}\left(  D^{\left(  k\right)  }\right)
\sigma^{P}\left(  D^{\left(  m\right)  }\right)  =\sigma^{P}\left(  D^{\left(
m\right)  }\right)  \sigma^{P}\left(  D^{\left(  k\right)  }\right)  $,
 as well as that:%
\begin{equation}
\left[  D_{1},D_{2}\cdot D_{3}\right]  =\left[  D_{1},D_{2}\right]  \cdot
D_{3}+D_{2}\cdot\left[  D_{1},D_{3}\right]
\end{equation}
for any three differential operators $D_{1},D_{2}$ and $D_{3}$.

It is possible to define a Lie algebra product on principal symbols by
associating with any two of them the principal symbol of their commutator. If
we denote by: $\sigma^{P}:Diff^{\left(  k\right)  }\left(  Q\right)
\rightarrow$ $\mathcal{S}^{\left(  k\right)  }\left(  Q\right)  $ the map that
associates a symbol with an operator, we can define:%
\begin{equation}
\left\{  \sigma^{P}\left(  D_{1}\right)  ,\sigma^{P}\left(  D_{2}\right)
\right\}  =:\sigma^{P}\left(  \left[  D_{1},D_{2}\right]  \right)
\label{PBPS}%
\end{equation}
In this way we define a Poisson bracket on the commutative algebra of the
principal symbols of differential operators. Moreover, we have:%
\begin{equation}
\left\{  \sigma^{P}\left(  D_{1}\right)  ,\sigma^{P}\left(  D_{2}\right)
\sigma^{P}\left(  D_{3}\right)  \right\}  =\sigma^{P}\left(  \left[
D_{1},D_{2}\right]  \right)  \sigma^{P}\left(  D_{3}\right)  +\sigma
^{P}\left(  D_{2}\right)  \sigma^{P}\left(  \left[  D_{1},D_{3}\right]
\right)
\end{equation}
which shows that \ $\left\{  \sigma^{P}\left(  D\right)  ,\cdot\right\}  $ is
a derivation on the commutative algebra of the principal symbols.

It is not difficult to see from Eq.(\ref{PS2}) that one can define (and in an
unique way) a symmetric contravariant tensor $\mathcal{D}^{\left(  k\right)
}$ of rank $k$ via:%
\begin{equation}
\sigma^{P}\left(  D^{\left(  k\right)  }\right)  \left(  \mathbf{f}\right)
=\mathcal{D}^{\left(  k\right)  }\left(  df_{1},df_{2},...,df_{k}\right)
\label{PS3}%
\end{equation}
evaluated on the symmetrized product $df_{1}\otimes df_{2}\otimes...\otimes
df_{k}$. \\
For example, for a homogeneous second-order operator of the form:%
\begin{equation}
D^{\left(  2\right)  }=\frac{1}{2}a^{ij}\frac{\partial^{2}}{\partial
q^{i}\partial q^{j}}%
\end{equation}
$\left(  a^{ij}=a^{ji}\right)  $ we find:%
\begin{equation}
\mathcal{D}^{\left(  2\right)  }=\frac{1}{2}a^{ij}\frac{\partial}{\partial
q^{i}}\otimes\frac{\partial}{\partial q^{j}}\label{tens}%
\end{equation}\\
This characterization of principal symbols by means of symmetric
contravariant tensor fields allows us to conclude that the Poisson bracket on
principal symbols is isomorphic with the Lie algebra product on symmetric
contravariant tensors defined by the Schouten bracket \cite{Sch}.
\begin{remark}
 If $Q$\textit is parallelizable, there will be a
global basis, say $\left(  X_{1},...,X_{n}\right) $ of vector
fields, and we can consider the $\mathcal{F}\left(  Q\right)  $-module of
contravariant tensor fields generated by them. For example, the monomial
$X_{i_{1}}\otimes...\otimes X_{i_{k}}$ defines a differential
operator of order $k$ by setting:%
\begin{equation}
D^{\left(  k\right)  }\left(  f\right)  =L_{X_{i_{1}}}\left(  L_{X_{i_{2}}%
}...\left(  L_{X_{i_{k}}}f\right)  \right)
\end{equation}
which corresponds to the differential operator $\underset{k\text{
}commutators}{\underbrace{\left[  ...\left[  D^{\left(  k\right)  }%
,\widehat{f}\right]  ...\widehat{f}\right]  }}$, which is of order
zero.
\end{remark}

A further identification is possible by considering the principal symbol
$\sigma^{P}\left(  D^{\left(  k\right)  }\right)  $ as a fiberwise polynomial
function $f_{D^{\left(  k\right)  }}$ on $T^{\ast}Q$ defined as:%
\begin{equation}
f_{D^{\left(  k\right)  }}=:\mathcal{D}^{\left(  k\right)  }\left(  \theta
_{0}\otimes\theta_{0}\otimes...\otimes\theta_{0}\right)  \label{PS4}%
\end{equation}\\
For example, with: $\theta_{0}=p_{i}dq^{i}$, the rank-two tensor (\ref{tens})
leads to:%
\begin{equation}
f_{D^{\left(  2\right)  }}=\frac{1}{2}a^{ij}p_{i}p_{j}%
\end{equation}\\
Let us stress the fact that what Eqs.(\ref{PS3}) and (\ref{PS4}) show is that
it is possible to characterize principal symbols completely in
\textit{tensorial} terms.

\begin{remark}
We are committing a slight abuse of notation in
making this definition, as we are contracting tensor fields on $Q$
with forms on $T^{\ast}Q$. However, the abuse is justified by the
fact that $\theta_{0}$ is a semi-basic one-form.
\end{remark}
With this association, the Poisson bracket (associated with $d\theta_{0}$) on
polynomial functions defined by contracting $\mathcal{D}^{\left(  k\right)  }$
with covariant tensor fields defined by means of powers of $\theta_{0}$ turns
out to be isomorphic with the "abstract" Poisson bracket (\ref{PBPS}) defined
on principal symbols of differential operators.

We should stress here that this realization in terms of functions that are
polynomials in the momenta depends on the semi-basic one-form we have used
(even though $\theta_{0}$ is a natural one-form \cite{MSSV} on $T^{\ast}Q$).
It would be possible to consider other semi-basic one-forms whose exterior
derivative would be a symplectic structure.\\
For example, we might consider, for any non-singular numerical matrix
$K=\left\Vert K^{i}\text{ }_{j}\right\Vert $:
\begin{equation}
\theta_{K}=p_{i}K^{i}\text{ }_{j}dq^{j}\label{K-form}%
\end{equation}
whereby: $d\theta_{K}=dp_{i}K^{i}$ $_{j}\wedge dq^{j}$ will be a symplectic
structure. Then, again using the rank-two tensor (\ref{tens}), one finds the quadratic form:%
\begin{equation}
\mathcal{D}^{\left(  2\right)  }\left(  \theta_{K}\otimes\theta_{K}\right)
=\frac{1}{2}p_{i}\left(  Ka^{t}K\right)  ^{ij}p_{j}%
\end{equation}
$\left(  \left(  ^{t}K\right)  _{j}\text{ }^{i}=K^{i}\text{ }_{j}\right)  $.
In this way, again the associated
Poisson algebra would be isomorphic with the abstract Poisson algebra defined
by means of the principal symbols. These alternatives may turn out to be
relevant if we would like to consider bi-Hamiltonian systems. In such a
situation, semi-basic one-forms $\widetilde{\theta}$ such that:%
\begin{equation}
L_{\Gamma}\widetilde{\theta}=d\widetilde{F}%
\end{equation}
which implies:%
\begin{equation}
i_{\Gamma}d\widetilde{\theta}=-d\widetilde{H};\text{ }\widetilde{H}=:i_{\Gamma}\widetilde{\theta}-\widetilde{F}
\end{equation}
would be of particular interest.

\textbf{Example 5.} \ \textit{On }$T^{\ast}\mathbb{R}^{2}$ \textit{with
coordinates }$\left(  q^{1},q^{2},p_{1},p_{2}\right)  $ \textit{the dynamics
of the }$2D$ \textit{isotropic harmonic oscillator can be represented by the
vector field:}%
\begin{equation}
\Gamma=p_{1}\frac{\partial}{\partial q^{1}}+p_{2}\frac{\partial}{\partial
q^{2}}-q^{1}\frac{\partial}{\partial p_{1}}-q^{2}\frac{\partial}{\partial
p_{2}}%
\end{equation}
\textit{which is Hamiltonian w.r.t. the "canonical" symplectic form }%
$\omega_{0}=d\theta_{0}=\sum_{i}dp_{i}\wedge dq^{i}$ \textit{with the
"standard" Hamiltonian: }$H=\sum_{i}(\left(  p_{i}\right)  ^{2}+\left(
q^{i}\right)  ^{2})/2$. \textit{It is also Hamiltonian  (and hence bi-Hamiltonian) w.r.t. (among others \cite{MFLMR}) the symplectic form:}%
\begin{equation}
\widetilde{\omega}=d\widetilde{\theta};\text{ }\widetilde{\theta}=p_{2}%
dq^{1}+p_{1}dq^{2}%
\end{equation}
\textit{which is of the form (\ref{K-form}) with: }%
\begin{equation}
K=\left\vert
\begin{array}
[c]{cc}%
0 & 1\\
1 & 0
\end{array}
\right\vert
\end{equation}
\textit{and where:}%
\begin{equation}
L_{\Gamma}\widetilde{\theta}=d\widetilde{F};\text{ }\widetilde{F}=p_{1}%
p_{2}-q^{1}q^{2}%
\end{equation}

\bigskip

\subsection{Hamilton-Jacobi-Type Equations Associated with Symbols.\label{HJS}%
}

Up to now we have seen that with the principal symbol of any differential
operator we can associate both a symmetric contravariant tensor field and a
polynomial function on $T^{\ast}Q$. Clearly, if we consider this polynomial
function $f_{D}$, it is possible to construct a first-order $PDE$ by setting,
as in Sect.\ref{TIHJ}:%
\begin{equation}
\left(  dS\right)  ^{\ast}\left(  f_{D}\right)  =0 \label{HJ13}%
\end{equation}
or, more generally:%
\begin{equation}
\left(  dS\right)  ^{\ast}\left(  f_{D}\right)  =c \label{HJ14}%
\end{equation}
thus defining a $PDE$ of the Hamilton-Jacobi type.

>From what we have said, we may set, equivalently:%
\begin{equation}
\left[  ...\left[  D^{\left(  k\right)  },\widehat{S}\right]  ,...,\widehat
{S}\right]  =\widehat{c} \label{HJ15}%
\end{equation}
so that we can define the $HJ$ equation associated with $D^{\left(  k\right)
}$ directly on the configuration manifold $Q$, without making recourse to the
cotangent bundle.

Out of the cotangent bundle representation of the principal symbol $\sigma
^{P}\left(  D\right)  $, i.e. by means of the polynomial functions $f_{D}$ on
$T^{\ast}Q$, it is possible to associate also Hamilton's equations constructed
out of $f_{D}$, namely:%
\begin{equation}
i_{\Gamma}d\theta_{0}=-df_{D}=-d\left( \mathcal{D}\left(  \theta_{0}%
\otimes\theta_{0}\otimes...\otimes\theta_{0}\right)  \right)  \label{Ham}%
\end{equation}

In this way, by means of the principal symbols, we are able to construct a
first-order $PDE$ of the Hamilton-Jacobi type as well as Hamilton's equations
that define the vector field $\Gamma$. Let us look now at a couple
of examples.

\textbf{Example 6.} \textit{The Schr\"{o}dinger operator.}
\textit{The operator associated with the evolution equation is:}%
\begin{equation}
D_{S}=i\hbar\frac{\partial}{\partial t}+\frac{\hbar^{2}}{2m}\nabla
^{2}-V\left(  \mathbf{r}\right)
\end{equation}

\textit{The associated }$HJ$ \textit{equation may be written in the form:}%
\begin{equation}
\left[  \left[  D_{S},\widehat{S}\right]  ,\widehat{S}\right]  =E
\end{equation}
\textit{i.e.:}%
\begin{equation}
\frac{\hbar^{2}}{2m}\left(  \nabla S\right)  ^{2}=E
\end{equation}

\textit{We find that the principal symbol, or the }$HJ$ \textit{equation
associated with it, contains no information on the potential nor on the time
evolution.}

\bigskip

\textbf{Example 7. }\textit{The Klein-Gordon operator.}
\textit{A similar situation prevails for the Klein-Gordon operator:}%
\begin{equation}
D_{KG}=\nabla^{2}-\frac{1}{c^{2}}\frac{\partial^{2}}{\partial t^{2}}-m^{2}%
\end{equation}
\textit{which leads to:}%
\begin{equation}
\left(  \nabla S\right)  ^{2}-\frac{1}{c^{2}}\left(  \frac{\partial
S}{\partial t}\right)  ^{2}=0
\end{equation}
\textit{i.e. the principal symbol and the associated }$HJ$ \textit{equation do
not take into account the mass of the particle.}

We see from these examples that the principal symbol does not capture the full
physical information contained in the differential operator (with the
associated $PDE$) we started from. To remedy this situation, we may define an
associated, homogeneous, differential operator on a larger space by adding one
more degree of freedom. To be specific, let us consider the case of a
second-order differential operator. Locally, the operator will have the
representation:%
\begin{equation}
D=a_{jk}\frac{\partial}{\partial x_{j}}\frac{\partial}{\partial x_{k}}%
+b_{j}\frac{\partial}{\partial x_{j}}+e
\end{equation}
($a_{ij},b_{j},e\in\mathcal{F}\left(  Q\right)  $). Adding then one more
variable, denoted as $\tau$, we obtain the extended differential operator on
$Q\times\mathbb{R}$:%

\begin{equation}
D\rightarrow\widetilde{D}=a_{jk}\frac{\partial}{\partial x_{j}}\frac{\partial
}{\partial x_{k}}+b_{j}\frac{\partial}{\partial x_{j}}\frac{\partial}%
{\partial\tau}+e\frac{\partial^{2}}{\partial\tau^{2}}%
\end{equation}
which is now homogeneous of degree two. We will also restrict the space of functions on which $\widetilde{D}$ operates
to functions of the form:%
\begin{equation}
\widetilde{f}\left(  x,\tau\right)  =e^{\tau}f\left(  x\right)
\end{equation}
in such a way that:%
\begin{equation}
\widetilde{D}\left(  \widetilde{f}\right)  \equiv e^{\tau}D\left(  f\right)
\end{equation}
The polynomial function on $T^{\ast
}\left(  Q\times\mathbb{R}\right)  $ (with coordinates: $\left(  x_{j}%
,\tau;p^{j},p^{\tau}\right)  $) associated with the principal symbol of
$\widetilde{D}$ will be:%
\begin{equation}
f_{\widetilde{D}}=a_{jk}p^{j}p^{k}+b_{j}p^{j}p^{\tau}+e\left(
p^{\tau
}\right)  ^{2}%
\end{equation}
and we will look for solutions $HJ$ equation:%
\begin{equation}
\left(  d\widetilde{S}\right)  ^{\ast}\left(  f_{\widetilde{D}}\right)
=c\label{HJ16}%
\end{equation}
of the form:%
\begin{equation}\label{sol}
\widetilde{S}\left(  x,\tau\right)  =\tau+S\left(  x\right)
\end{equation}
In this way, the \ $HJ$ equation will become, when written in local
coordinates:%
\begin{equation}
a_{jk}\frac{\partial S}{\partial x_{j}}\frac{\partial S}{\partial x_{k}}%
+b_{j}\frac{\partial S}{\partial x_{j}}+e=c
\end{equation}
and we have gotten rid in this way of the additional degree of freedom whose
introduction was made necessary in order to be able to deal with tensorial
objects (see the discussion in Sect.\ref{DOPS}).

\bigskip

\textbf{Example 8.} \textit{With the procedure outlined above, both the
Schr\"{o}dinger and the Klein-Gordon operators get replaced by:}%
\begin{equation}
D_{S}\rightarrow\widetilde{D}_{S}=i\hbar\frac{\partial^{2}}{\partial
t\partial\tau}+\frac{\hbar^{2}}{2m}\nabla^{2}-V\left(  \mathbf{r}\right)
\frac{\partial^{2}}{\partial\tau^{2}}%
\end{equation}
\textit{and:}%
\begin{equation}
D_{KG}\rightarrow\widetilde{D}_{KG}=\nabla^{2}-\frac{1}{c^{2}}\frac
{\partial^{2}}{\partial t^{2}}-m^{2}\frac{\partial^{2}}{\partial\tau^{2}}%
\end{equation}
\textit{respectively, and the associated }$HJ$ \textit{equations (looking for
solutions of the form (\ref{sol})) will capture the full physics of the
respective problems.}

The procedure outlined here is completely general, and we see that in this way the full
information contained in our original differential operator will be captured
by the principal symbol of the extended operator, and we can proceed now with
the first-order $PDE$ and Hamilton's equations just as before. As for transformations, we should remark that now we have to restrict to bundle automorphisms.

\bigskip

\subsection{Vector-Valued Differential Operators and the Hamilton-Jacobi Problem.\label{VVDO}}

In several physical situations, the systems we may want to describe have also
some inner structure, and therefore scalar differential operators are not
enough. For instance, the quantum-mechanical description of a particle with
spin requires that we replace Schr\"{o}dinger's equation with Pauli's in the
non-relativistic case and with Dirac' in the relativistic case. More
generally, this kind of situation occurs when we consider \cite{BMSS1,BMSS2} Yang-Mills fields
and generalized Wong equations.

Let us consider then the general aspects of this situation. We consider here
two vector bundles: $E_{1}\rightarrow Q$ and: $E_{2}\rightarrow Q$. We denote
by $Sec\left(  E_{1}\right)  $ and $Sec\left(  E_{2}\right)  $ the spaces of
sections: $s:Q\rightarrow E_{1}$ and: $\tau:Q\rightarrow E_{2}$. The operator
of multiplication by a function $f\in\mathcal{F}\left(  Q\right)  $ will be
denoted here too by $\widehat{f}$. Following the algebraic setting for
differential operators of Sect.\ref{DOPS}, we define:

\textit{A differential operator of order at most }$k$\textit{, acting from
}$Sec\left(  E_{1}\right)  $\textit{ to }$Sec\left(  E_{2}\right)  $\textit{,
is a linear map:}%
\begin{equation}
D:Sec\left(  E_{1}\right)  \longrightarrow Sec\left(  E_{2}\right)
\end{equation}
\textit{such that:}%
\begin{equation}
\left[  ...\left[  \left[  D^{\left(  k\right)  },\widehat{f}_{0}\right]
,\widehat{f}_{1}\right]  ,...,\widehat{f}_{k}\right]  =0\text{ \ }%
\forall\widehat{f}_{0},\widehat{f}_{1},...,\widehat{f}_{k}%
\end{equation}

In a chosen trivialization of the two bundles, their sections are
vector-valued functions on $Q$, and the operator $D$ will be described by a
matrix whose entries are coordinate expressions of scalar differential
operators. Thus, here too the relevant differential part can be written as:%
\begin{equation}%
\begin{array}
[c]{c}%
D^{\left(  k\right)  }=%
{\displaystyle\sum\limits_{\left\vert \sigma\right\vert \leq k}}
g_{\sigma}\frac{\partial^{\left\vert \sigma\right\vert }}{\partial x_{\sigma}%
}\\
\sigma=\left(  i_{1},i_{2},...,i_{n}\right)  ;\text{ }\left\vert
\sigma\right\vert =i_{1}+i_{2}+...+i_{n}\\
\partial^{\left\vert \sigma\right\vert }/\partial x_{\sigma}=\partial
^{\left\vert \sigma\right\vert }/\partial x_{1}^{i_{1}}\partial x_{2}^{i_{2}%
}...\partial x_{n}^{i_{n}}%
\end{array}
\end{equation}
but now, instead of the $g_{\sigma}$'s being pointwise scalars, we have:%
\begin{equation}
g_{\sigma}\in Hom\left(  E_{1},E_{2}\right)
\end{equation}
(pointwise), and each matrix element of the $g_{\sigma}$'s will be \ a smooth
function on $Q$.

We can again identify the principal symbol with a symmetric and totally
contravariant tensor, i.e.:%
\begin{equation}
\sigma^{P}\left(  D^{\left(  k\right)  }\right)  =g_{i_{1}i_{2}...i_{k}}%
\frac{\partial}{\partial x_{i_{1}}}\otimes\frac{\partial}{\partial x_{i_{2}}%
}\otimes...\otimes\frac{\partial}{\partial x_{i_{k}}}%
\end{equation}
but now the tensor field must be understood as a multilinear function on
one-forms with values in $Hom\left(  E_{1},E_{2}\right)  $. Intrinsically:%
\begin{equation}
\sigma^{P}\left(  D^{\left(  k\right)  }\right)  \left(  df_{1},df_{2}%
,...,df_{k}\right)  =\left[  ...\left[  \left[  D^{\left(  k\right)
},\widehat{f}_{k}\right]  ,...\widehat{f}_{2}\right]  ,\widehat{f}_{1}\right]
\end{equation}
and the symmetry follows again from Eq.(\ref{symm}). By using the contraction
with the $k$-fold product of $\theta_{0}$ with itself we obtain a polynomial
function with matrix-valued coefficients.

\textbf{Example 9.} \textit{Let; }$D=d$\textit{, the exterior differential:
}$d:\Lambda^{j}\left(  Q\right)  \rightarrow\Lambda^{j+1}\left(  Q\right)
$\textit{. The value of its symbol on a differential one-form }$\alpha
$\textit{ is a homomorphism from }$\Lambda^{j}\left(  Q\right)  $\textit{ to
}$\Lambda^{j+1}\left(  Q\right)  $\textit{ given by \cite{EMS}:}%
\begin{equation}
\sigma^{P}\left(  d\right)  \left(  \alpha\right)  :\Lambda^{j}\left(
Q\right)  \ni\omega\longmapsto\alpha\wedge\omega\in\Lambda^{j+1}\left(
Q\right)
\end{equation}

The symbol of second-order scalar differential operators are symmetric
contravariant tensor fields of rank two, that is:%
\begin{equation}
\sigma^{P}\left(  D\right)  =a_{jk}\frac{\partial}{\partial x_{j}}\otimes
\frac{\partial}{\partial x_{k}};\text{ \ }a_{jk}=a_{kj}%
\end{equation}

When this tensor is not degenerate, it defines a (pseudo) Riemannian metric:%
\begin{equation}
g_{D}=a^{jk}dx_{j}\otimes dx_{k}%
\end{equation}
with:%
\begin{equation}
a^{jk}a_{km}=\delta_{m}^{j}%
\end{equation}
Thus, $D$ is elliptic if $g_{D}$ is Riemannian and hyperbolic if it is Lorentzian.

In the scalar situation we have defined a Hamiltonian as:%
\begin{equation}%
\begin{array}
[c]{c}%
H\left(  q,p\right)  =\sigma^{P}\left(  D\right)  \left(  \theta_{0}%
\otimes...\otimes\theta_{0}\right)  =\\
=a_{i_{1}...i_{k}}p^{i_{1}}...p^{i_{k}}%
\end{array}
\end{equation}
and in this way we have been able to define Poisson brackets,
Hamilton-Jacobi-type equations and, finally, Hamilton's canonical equations.

In the present context, the symbol of a matrix-valued \ differential operator
$D$ will be a matrix-valued polynomial on $T^{\ast}Q$. For example, for a
second-order operator we will get:%
\begin{equation}
\sigma^{P}\left(  D\right)  \left(  \theta_{0}\otimes\theta_{0}\right)
=H\left(  q,p\right)
\end{equation}
where now $H\left(  q,p\right)  $ will be a matrix.

If our bundles were Hermitian bundles of the same dimension, they can be
identified and $H$ becomes Hermitian: $H=H^{\dag}$. Then, $H$ will have real
eigenvalues, and each eigenvalue function can be used as a Hamiltonian
function on $T^{\ast}Q$ \cite{Avra}. At this point we can repeat whatever has been said in
the case of scalar differential operators.

\bigskip

\bigskip

\bigskip

\bigskip

\bigskip

\section{Joint Hamilton-Jacobi Problems.}\label{joint0}
As anticipated in the Introduction, one can pose a "joint Hamilton-Jacobi"
problem also for two or more dynamical variables. For reasons that will become
apparent shortly, we will limit ourselves here to the "conventional" $HJ$
problem as discussed in Sect.\ref{TIHJ}. In order to discuss the "joint"
problem, we will have to discuss first some preliminary notions related to the
restriction of Poisson brackets (on a cotangent bundle) to Lagrangian
submanifolds that are graphs of closed one-forms.

In Sect.\ref{TIHJ} we have shown that, given the cotangent bundle $T^{\ast}Q$
$\ $of a configuration manifold $Q$ with the canonical symplectic form
$\omega_{0}$, with every one-form $\alpha\in\mathfrak{X}^{\ast}\left(
Q\right)  $ we can associate the maps (\ref{map1}) and (\ref{map2}), i.e. (we
omit here for brevity the suffix "$\alpha$" that was employed in
Sect.\ref{TIHJ}): \
\begin{equation}
\varphi:Q\rightarrow T^{\ast}Q\text{ }by:q\mapsto\left(  q,\alpha\left(
q\right)  \right)  \label{map3}%
\end{equation}
and:%
\begin{equation}
\psi=:\varphi\circ\pi:T^{\ast}Q\rightarrow T^{\ast}Q\text{ }by:\left(
q,p\right)  \mapsto\left(  q,\alpha\left(  q\right)  \right)  \label{map4}%
\end{equation}
and that the \textit{image} of $\alpha$: $\Gamma\left[  \alpha\right]
=\left(  \left(  q,\alpha\left(  q\right)  \right)  \in T^{\ast}Q;q\in
Q\right)  $ will be a Lagrangian submanifold of $T^{\ast}Q$ iff $\alpha$ is
closed. Notice that, if $\left(  q,p\right)  $ belongs already to
$\Gamma\left[  \alpha\right]  $, the map (\ref{map4}) will leave it unaltered:%
\begin{equation}
\psi\left(  \Gamma\left[  \alpha\right]  \right)  =\Gamma\left[
\alpha\right]
\end{equation}
in a fiber-preserving way.

With every function $f\in\mathcal{F}\left(  T^{\ast}Q\right)  $ we can
associate the pull-back:%
\begin{equation}
\left(  \psi^{\ast}f\right)  \left(  q,p\right)  =f\left(  q,\alpha\left(
q\right)  \right)
\end{equation}
i.e. the extension of the restriction of $f$ to $\Gamma\left[  \alpha\right]
$ with the property of being constant along the fibers. With this in mind, we
want here to relate the restriction $\psi^{\ast}\left\{  f,g\right\}  $ of the
Poisson bracket of any two functions $f$ and $g$ to $\Gamma\left[
\alpha\right]  $ to the corresponding restrictions $\psi^{\ast}f$ and
$\psi^{\ast}g$ of $f$ and $g$.

If $\pi:T^{\ast}Q\rightarrow Q$ is the canonical projection, Eq.(\ref{map4})
implies: $\pi\circ\psi=\pi$, and hence:
\begin{equation}
T\pi\circ T\psi=T\pi\label{tang1}%
\end{equation}
for the corresponding tangent maps.

Let now $m=\left(  q,p\right)  \in T^{\ast}Q$ and: $\gamma=\psi\left(
m\right)  $. Then to every tangent vector $X\in T_{m}T^{\ast}Q$ one can
associate the vector: $T\psi\left(  X\right)  \in T_{\gamma}T^{\ast}Q$ and
Eq.(\ref{tang1}) implies:%
\begin{equation}
T\pi\left(  T\psi\left(  X\right)  \right)  -T\pi\left(  X\right)  =0
\label{tang2}%
\end{equation}
Notice that the two arguments of $T\pi$ in Eq.(\ref{tang2}) are in general
tangent vectors at \textit{different} points of $T^{\ast}Q$. It is only when
$m=\gamma\in\Gamma\left[  \alpha\right]  $ that we can factor out the tangent
map $T\pi$ and write:%
\begin{equation}
T\pi\left(  \left(  T\psi\left(  X\right)  \right)  -X\right)  =0,\text{
\ }X\in T_{\gamma}T^{\ast}Q,\text{ }\gamma\in\Gamma\left[  \alpha\right]
\label{tang4}%
\end{equation}
What Eq.(\ref{tang4}) proves is that, if $X\in T_{\gamma}T^{\ast}Q$, then
$\left(  T\psi\left(  X\right)  \right)  -X$ is a \textit{vertical} field.

Let now $f,g\in\mathcal{F}\left(  T^{\ast}Q\right)  $ and let $X_{f},X_{g}$ be
the corresponding Hamiltonian vector fields defined via:%
\begin{equation}
i_{X_{f}}\omega_{0}=-df,\text{ }i_{X_{g}}\omega_{0}=-dg\text{\ }\label{Ham}%
\end{equation}
while:%
\begin{equation}
\left\{  f,g\right\}  =\omega_{0}\left(  X_{g},X_{f}\right)
\end{equation}
Then:%
\begin{equation}%
\begin{array}
[c]{c}%
\left(  \psi^{\ast}\left\{  f,g\right\}  \right)  \left(  m\right)  =\left\{
f,g\right\}  \left(  \gamma\right)  =\omega_{0}\left(  \gamma\right)  \left(
X_{g}\left(  \gamma\right)  ,X_{f}\left(  \gamma\right)  \right)  \equiv\\
\omega_{0}\left(  \gamma\right)  \left(  X_{g}\left(  \gamma\right)
-T\psi\left(  X_{g}\left(  \gamma\right)  \right)  ,X_{f}\left(
\gamma\right)  -T\psi\left(  X_{f}\left(  \gamma\right)  \right)  \right)  +\\
\omega_{0}\left(  \gamma\right)  \left(  X_{g}\left(  \gamma\right)
,T\psi\left(  X_{f}\left(  \gamma\right)  \right)  \right)  +\omega_{0}\left(
\gamma\right)  \left(  T\psi\left(  X_{g}\left(  \gamma\right)  \right)
,X_{f}\left(  \gamma\right)  \right)  -\\
\omega_{0}\left(  \gamma\right)  \left(  T\psi\left(  X_{g}\left(
\gamma\right)  \right)  ,T\psi\left(  X_{f}\left(  \gamma\right)  \right)
\right)
\end{array}
\end{equation}
Now, in the first term on the r.h.s. of this equation, $\omega_{0}$ is
evaluated on a pair of vertical fields (at $\gamma$), and hence this term
vanishes. As to the last term, using the definition of the pull-back
\cite{MSSV}:%
\begin{equation}
\omega_{0}\left(  \gamma\right)  \left(  T\psi\left(  X_{g}\left(
\gamma\right)  \right)  ,T\psi\left(  X_{f}\left(  \gamma\right)  \right)
\right)  =\left(  \psi^{\ast}\omega_{0}\right)  \left(  \gamma\right)  \left(
X_{g}\left(  \gamma\right)  ,X_{f}\left(  \gamma\right)  \right)
\end{equation}
But: $\psi^{\ast}\omega_{0}=d\alpha$ and, as $\alpha$ is closed by assumption,
this term vanishes as well and, using also Eq.(\ref{Ham}), we are left with:%
\begin{equation}%
\begin{array}
[c]{c}%
\left(  \psi^{\ast}\left\{  f,g\right\}  \right)  \left(  m\right)  =\left(
i_{T\psi\left(  X_{f}\left(  \gamma\right)  \right)  }i_{X_{g}\left(
\gamma\right)  }-i_{T\psi\left(  X_{g}\left(  \gamma\right)  \right)
}i_{X_{f}\left(  \gamma\right)  }\right)  \omega_{0}\left(  \gamma\right)  =\\
\left(  df\right)  \left(  \gamma\right)  \left(  T\psi\left(  X_{g}\left(
\gamma\right)  \right)  \right)  -(dg)\left(  \gamma\right)  \left(
T\psi\left(  X_{f}\left(  \gamma\right)  \right)  \right)  =\\
\left(  d\psi^{\ast}f\right)  \left(  X_{g}\left(  \gamma\right)  \right)
-\left(  d\psi^{\ast}g\right)  \left(  X_{g}\left(  \gamma\right)  \right)
=\\
\left(  L_{X_{g}}\psi^{\ast}f-L_{X_{f}}\psi^{\ast}g\right)  \left(
\gamma\right)
\end{array}
\end{equation}
The last term here is in turn the pull-back via $\psi$ of the function:
$L_{X_{g}}\psi^{\ast}f-L_{X_{f}}\psi^{\ast}g$, and in this way we obtain the
result \cite{MMN}:%
\begin{equation}
\psi^{\ast}\left\{  f,g\right\}  =\psi^{\ast}\left(  L_{X_{g}}\psi^{\ast
}f-L_{X_{f}}\psi^{\ast}g\right)  \label{restr}%
\end{equation}
which expresses the relation between the pull-back (the restriction) of the
Poisson bracket of any two functions and the restrictions of the functions
themselves.  All that has been proved here relies in a crucial way on
$\Gamma\left[  \alpha\right]  $ being the graph in a cotangent bundle of a
closed one-form on the base manifold, and cannot be extended
straightforwardly  \cite{MMN} to more general contexts such as Lagrangian
sumbanifolds in general symplectic manifolds without further qualifications
(see however Ref.\cite{MMN} for some possible generalizations).

Having established the result (\ref{restr}), let's turn now to what we have
called at the beginning of this Section the "joint $HJ$ problem". To be
specific, let: $A,B\in\mathcal{F}\left(  T^{\ast}Q\right)  $ be two dynamical
variables. We look then for a closed one-form $\alpha$ such that (cfr.
Eqs.(\ref{map3}) and (\ref{map4})):%
\begin{equation}
\psi^{\ast}\left(  A-a\right)  =\psi^{\ast}\left(  B-b\right)  =0;\text{
}a,b\in\mathbb{R}%
\end{equation}
or:%
\begin{equation}
\psi^{\ast}A=a,\text{ \ }\psi^{\ast}B=b
\end{equation}
Eq.(\ref{restr}) tells us immediately that this implies:%
\begin{equation}
\psi^{\ast}\left\{  A,B\right\}  =\psi^{\ast}\left(  L_{X_{B}}\psi^{\ast
}A-L_{X_{A}}\psi^{\ast}B\right)  =0
\end{equation}
This (\textit{necessary}) condition, i.e.:%
\begin{equation}
\psi^{\ast}\left\{  A,B\right\}  =0\label{joint}%
\end{equation}
can be interpreted \cite{MMN} as a sort of Poisson theorem, i.e. a
condition saying that if a solution exists for the joint $HJ$
problem for $A$ and $B$, then it must be also a solution for the
$HJ$ problem for the Poisson bracket $\left\{  A,B\right\}  $ on its
zero level-set, and also as a classical counterpart of the quantum
condition \cite{Me} according to which two observables must commute
in order to be simultaneously diagonalizable. It is obvious that if
the Poisson bracket $\left\{  A,B\right\}  $ does not vanish
anywhere, then there is no possible solution for the joint $HJ$
problem for $A$ and $B$.

The obvious generalization to any set $A_{1},...,A_{k},k\leq n=\dim\left(
Q\right)  $ will be, of course:%
\begin{equation}
\psi^{\ast}\left\{  A_{i},A_{j}\right\}  =0;\text{ }i,j=1,...,k
\end{equation}

As the one-form $\alpha$ is closed by assumption, at least locally
(and globally if the first de Rham cohomology group of $Q$
vanishes): $\alpha=dW,W\in\mathcal{F}\left( Q\right) $, and
Eq.(\ref{joint}) will become a $PDE$ for the unknown function $W$.
For example, in local (Darboux) coordinates,
$\omega_{0}=dp_{i}\wedge
dq^{i}$ will imply the equation:%
\begin{equation}
\left[  \frac{\partial A\left(  q,p\right)  }{\partial q^{i}}\frac{\partial
B\left(  q,p\right)  }{\partial p_{i}}-\frac{\partial B\left(  q,p\right)
}{\partial q^{i}}\frac{\partial A\left(  q,p\right)  }{\partial p_{i}}\right]
%
\genfrac{\vert}{.}{0pt}{}{{}}{p_{i}=\partial W/\partial q^{i}}%
=0\label{PDE}%
\end{equation}

We will close this Section by discussing a simple example that can help clarifying the status of the condition (\ref{joint}) as a necessary but not sufficient condition.

\textbf{Example 10.} \textit{It is well known \cite{MFLMR} that the dynamics of the
$2D$\textit{ harmonic oscillator on }$T^{\ast}R^{2}$}\textit{, with coordinates
}$\left(  q^{1},q^{2},p_{1},p_{2}\right)  $\textit{ and equipped with the
canonical symplectic form: }$\omega_{0}=dp_{j}\wedge dq^{j}$ \textit{can be
described by in many alternative ways and, among others, by:}

\begin{itemize}
\item \textit{The "standard" Hamiltonian:}%
\begin{equation}
H_{+}=H_{1}+H_{2};\text{ }H_{i}=\frac{1}{2}\left[  \left(  p_{i}\right)
^{2}+\left(  q^{i}\right)  ^{2}\right]  ,i=1,2\text{ }%
\end{equation}
\textit{and vector field:}%
\begin{equation}
\Gamma_{+}=\Gamma_{1}+\Gamma_{2};\text{ }\Gamma_{i}=p_{i}\frac{\partial
}{\partial q^{i}}-q^{i}\frac{\partial}{\partial p_{i}},i=1,2
\end{equation}
\textit{or by:}

\item \textit{The Hamiltonian and vector field:}%
\begin{equation}
H_{-}=H_{1}-H_{2};\text{ }\Gamma_{-}=\Gamma_{1}-\Gamma_{2}\label{H2}%
\end{equation}
\textit{or, eventually, by:}

\item
\begin{equation}
\widetilde{H}=p_{1}p_{2}+q^{1}q^{2}\label{Ham3}%
\end{equation}
\textit{and:}%
\begin{equation}
\widetilde{\Gamma}=p_{2}\frac{\partial}{\partial q^{1}}+p_{1}\frac{\partial
}{\partial q^{2}}-q^{2}\frac{\partial}{\partial p_{1}}-q^{1}\frac{\partial
}{\partial p_{2}}%
\end{equation}
\textit{which describes the dynamics of the $2D$ harmonic oscillator
on the configuration space.}
\end{itemize}

\textit{Now, it is clear that: }$\left\{  H_{+},H_{-}\right\}  \equiv
0$\textit{, each Hamiltonian being a constant of the motion for the other, and
the joint }$HJ$ \textit{problem for them is trivial, as complete integrals can
be found simply\ by separation of variables, i.e. by setting: }$W(q^{1}%
,q^{2})=W_{1}\left(  q^{1}\right)  +W_{2}\left(  q^{2}\right)  $,
\textit{thereby splitting the }$HJ$\textit{ problem for }$H_{\pm}$
\textit{into separate (and identical) one-dimensional }$HJ$ \textit{problems
for }$H_{1}$ \textit{and }$H_{2}$\textit{. Considering instead the
(alternative) Hamiltonian (\ref{Ham3}), we find again: }$\left\{
H_{+},\widetilde{H}\right\}  \equiv0$, but:%
\begin{equation}
\left\{  H_{-},\widetilde{H}\right\}  =2\left(  q^{1}p_{2}-q^{2}p_{1}\right)
\end{equation}
\textit{(i.e. the "canonical" angular momentum) and the $PDE$ equation (\ref{PDE})
becomes then}:%
\begin{equation}
q^{1}\frac{\partial W}{\partial q^{2}}-q^{2}\frac{\partial W}{\partial q^{1}%
}=0\label{PDE4}%
\end{equation}
\textit{whose only solutions\footnote{Switching to plane polar coordinates $\left(  q^{1}=r\cos\phi,q^{2}=r\sin
\phi\right)  :q^{1}\partial/\partial q^{2}-q^{2}\partial/\partial
q^{1}=\partial/\partial\phi$.} are of the form:}%
\begin{equation}
W\left(  q^{1},q^{2}\right)  =W\left(r^{2}\right)  ;\text{ }%
r^{2}=\left(  q^{1}\right)  ^{2}+\left(  q^{2}\right)  ^{2}\label{funct1}%
\end{equation}
\textit{with $W$ any (at least $C^{1}$) function. On the other hand, the $HJ$ equation
for the Hamiltonian $\widetilde{H}$ reads}:%
\begin{equation}
\frac{\partial W}{\partial q^{1}}\frac{\partial W}{\partial q^{2}}+q^{1}%
q^{2}=E\label{funct2}%
\end{equation}
\textit{and no function of the form (\ref{funct1})\footnote{Inserting (\ref{funct1})
into (\ref{funct2}) leads to the equation: $q^{1}q^{2}\left(  2(W^{\prime}%
)^{2}+1\right)  =E$, which has obviously no solution.} can solve
Eq.(\ref{funct2}), nor it can for the $HJ$ equation associated with the Hamiltonian (\ref{H2}). This clarifies the fact that the vanishing of the
restriction of the Poisson bracket is only a necessary and by no means a
sufficient condition for a solution of the joint $HJ$ problem}.

\section{Time-Dependent Hamilton-Jacobi Theory and the Hamilton-Jacobi Problem
on Lie Groups.}

\subsection{Time-Dependent Hamilton-Jacobi Theory.}\label{TDHJ_sec}

We will recall briefly here how the time-dependent $HJ$ theory as described,
in local coordinates, by Eq.(\ref{HJ1}), i.e.:%
\begin{equation}
H\left(  q;\frac{\partial S}{\partial q};t\right)  +\frac{\partial S}{\partial
t}=0 \label{time_dep}%
\end{equation}
can be recasted in a geometrical setting \cite{MMN} just as its
time-independent counterpart.

Although the "minimal" extension of the carrier manifold \cite{MSSV}
appropriate to the description of a time-dependent dynamics seems to be from
$T^{\ast}Q$ to $T^{\ast}Q\times\mathbb{R}$, where $\mathbb{R}$ stands for the
time variable $t$, the resulting manifold, being odd-dimensional, is a contact
\cite{AM} and not a symplectic manifold. As anticipated in Sect.\ref{comments}%
, it is more convenient \cite{AM,MMN,MSSV} to extend the original
configuration space $Q$ (with local coordinates $q^{i},i=1,...,n=\dim Q$) to:
$\widetilde{Q}=Q\times\mathbb{R}$, considering then time as an additional
coordinate on the same footing as the $q^{i}$'s, and hence to consider the
(extended) tangent bundle:%
\begin{equation}
\widetilde{\mathcal{M}}=:T^{\ast}\widetilde{Q}=T^{\ast}Q\times T^{\ast
}\mathbb{R}=\mathcal{M}\times T^{\ast}\mathbb{R};\text{ }\mathcal{M=}T^{\ast}Q
\end{equation}
Coordinates\footnote{\textit{Global} coordinates, in this case.} for $T^{\ast
}\mathbb{R}$ will be denoted as $\left(  t,h\right)  $, with $h$ the (energy)
variable canonically conjugate to the time $t$, and a point $\widetilde{m}\in$
$\widetilde{\mathcal{M}}$ will consist of a triple: $\widetilde{m}=\left(
m,t,h\right)  $ with $m\in\mathcal{M}$ (hence: $m=\left(  q,p\right)  $ in
local coordinates). Associated with the manifold $\widetilde{\mathcal{M}}$
there will be various projection maps, for instance from $T^{\ast}Q\times
T^{\ast}\mathbb{R}$ to $T^{\ast}Q\times\mathbb{R}$ or to $T^{\ast}Q$ or to
$T^{\ast}\mathbb{R}$ and so on. One can further endow $\widetilde{\mathcal{M}%
}$ with a canonical one-form $\widetilde{\theta}_{0}$ obtained by "adding" a
contribution from $T^{\ast}\mathbb{R}$ to $\theta_{0}=p_{i}dq^{i}$ on
$\mathcal{M}$, i.e.:%
\begin{equation}
\widetilde{\theta}_{0}=:\theta_{0}+hdt \label{can1}%
\end{equation}
Hence, $\widetilde{\mathcal{M}}$ will acquire the structure of a symplectic
manifold with the canonical two-form\footnote{To be precise, one should write
on the r.h.s. of Eqs.(\ref{can1}) and (\ref{can2}) $\pi^{\ast}\theta_{0}$ and
$\pi^{\ast}\omega_{0}$, where $\pi$ is the projection: $\pi:T^{\ast}Q\times
T^{\ast}\mathbb{R}\rightarrow T^{\ast}Q$, instead of $\theta_{0}$ and
$\omega_{0}$, but we will avoid, here and in what follows, explicit mention of
the pull-back operations involved as long as this does not lead to
ambiguities.}:%
\begin{equation}
\widetilde{\omega}_{0}=\omega_{0}+dh\wedge dt \label{can2}%
\end{equation}

The canonical equations of motion on $\mathcal{M}=T^{\ast}Q$ can be "lifted"
to $\widetilde{\mathcal{M}}=T^{\ast}\left(  Q\times\mathbb{R}\right)  $ by
defining, guided by the structure of the time-dependent equation
(\ref{time_dep}), the Hamiltonian $\widetilde{H}\in\mathcal{F(}\widetilde
{\mathcal{M}})$ on the extended phase space $\widetilde{\mathcal{M}}$ as:%
\begin{equation}
\widetilde{H}=H+h
\end{equation}
In this way:

\begin{itemize}
\item The Hamiltonian vector fields $\widetilde{X}_{H},\widetilde
{X}_{\widetilde{H}}\in\mathfrak{X}(\widetilde{M})$ associated respectively
with $H$ and $\widetilde{H}$ ($i_{\widetilde{X}_{H}}\widetilde{\omega}%
_{0}=-dH,i_{\widetilde{X}_{\widetilde{H}}}\widetilde{\omega}_{0}%
=-d\widetilde{H}$) are easily found to be:%
\begin{equation}
\widetilde{X}_{H}=X_{H}-\frac{\partial H}{\partial t}\frac{\partial}{\partial
h};\text{ }\widetilde{X}_{\widetilde{H}}=\widetilde{X}_{H}+\frac{\partial
}{\partial t} \label{can3}%
\end{equation}
($X_{H}=\left(  \partial H/\partial p\right)  \partial/\partial q$ $-\left(
\partial H/\partial q\right)  \partial/\partial p$), and hence:

\item Denoting with $\tau$ the evolution parameter in $\widetilde{\mathcal{M}%
}$, the dynamics generated by $\widetilde{X}_{\widetilde{H}}$ is described by
the set of $ODE$'s in the parameter $\tau$:%
\begin{equation}
\frac{dq}{d\tau}=\frac{\partial H}{\partial p},\text{ }\frac{dp}{d\tau}%
=-\frac{\partial H}{\partial q} \label{can4}%
\end{equation}
and:%
\begin{equation}
\frac{dt}{d\tau}=1,\text{ }\frac{dh}{d\tau}=-\frac{\partial H}{\partial t}
\label{can5}%
\end{equation}
These equations imply: $t-\tau=const.$ (and hence Eqs.(\ref{can4})
will "project down" to the canonical equations on $T^{\ast}Q$). They
imply also: $d\widetilde{H}/d\tau=0$, i.e.:
$\widetilde{H}=H+h=const.$

\item Denoting with $\left\{  .,.\right\}  $ and $\widetilde{\left\{
.,.\right\}  }$ the Poisson brackets on $\mathcal{M}=T^{\ast}Q$ and
$\widetilde{\mathcal{M}}=T^{\ast}(Q\times\mathbb{R})$ respectively, we have:%
\begin{equation}
\widetilde{\left\{  F,G\right\}  }=\left\{  F,G\right\}  +\frac{\partial
F}{\partial t}\frac{\partial G}{\partial h}-\frac{\partial F}{\partial h}%
\frac{\partial G}{\partial t},\text{ }\forall F,G\in\mathcal{F}\left(
\widetilde{\mathcal{M}}\right)
\end{equation}
and, in particular, if: $F\in\mathcal{F}\left(  T^{\ast}Q\times\mathbb{R}%
\right)  $ (i.e. it does not depend on $h$) and: $G=\widetilde{H}$:
\begin{equation}
\widetilde{\left\{  F,\widetilde{H}\right\}  }=\left\{  F,H\right\}
+\frac{\partial F}{\partial t}%
\end{equation}
i.e:%
\begin{equation}
\frac{dF}{dt}\equiv\widetilde{\left\{  F,\widetilde{H}\right\}  }%
\end{equation}
and (possibly time-dependent) constants of the motion will be characterized
by:%
\begin{equation}
\widetilde{\left\{  F,\widetilde{H}\right\}  }=0
\end{equation}
in full analogy with the time-independent case.
\end{itemize}

\bigskip

The dynamics on $\widetilde{\mathcal{M}}$ leaves invariant (i.e.
$\widetilde{X}_{\widetilde{H}}$ is tangent to) all the submanifolds where
$\widetilde{H}$ \ takes a constant value, and the dynamics on $T^{\ast}%
Q\times\mathbb{R}$ can be recovered from any one of them. For this reason, one
can concentrate on the zero-level set of $\widetilde{H}$, i.e. on the
invariant submanifold\footnote{Being of codimension one, $\Sigma_{0}$ will be
a coisotropic \cite{MMN} submanifold of $\widetilde{\mathcal{M}}$.}:%
\begin{equation}
\widetilde{\Sigma}_{0}=\widetilde{H}^{-1}\left(  0\right)  =\left(
m,t,h=-H\left(  m,t\right)  \right)  \subset\widetilde{\mathcal{M}};\text{
}m\in\mathcal{M},t\in\mathbb{R}%
\end{equation}
Being obtained by giving explicitly the (globally defined) variable $h$ in
terms of $m$ and $t$, $\widetilde{\Sigma}_{0}$ is diffeomorphic to $T^{\ast
}Q\times\mathbb{R}$, and is actually a global section of the projection map:
$\pi:T^{\ast}\left(  Q\times\mathbb{R}\right)  \rightarrow T^{\ast}%
Q\times\mathbb{R}$. \

If we keep $t$ fixed, we obtain a family $\{(\widetilde{\Sigma}_{0}%
)_{t}\}_{t\in\mathbb{R}}$ of constant-$t$ sections of $\widetilde{\Sigma}_{0}%
$. Each section is a submanifold of $T^{\ast}\left(  Q\times\mathbb{R}\right)
$ of codimension two (i.e. it is $2n$-dimensional, while $\widetilde{\Sigma
}_{0}$ is $\left(  2n+1\right)  $-dimensional) diffeomorphic to $T^{\ast}Q$
and is a global section of the projection map: $\pi^{\prime}:T^{\ast}\left(
Q\times\mathbb{R}\right)  \rightarrow T^{\ast}Q$. This family provides then a
(regular) foliation \cite{MSSV} of $\widetilde{\Sigma}_{0}$. Moreover, due to
the additional and nowhere vanishing $\partial/\partial t$ which is present in
the definition of $\widetilde{X}_{\widetilde{H}}$, the $(\widetilde{\Sigma
}_{0})_{t}$'s are transversal to the flow generated by $\widetilde
{X}_{\widetilde{H}}$, and the one-parameter group $\{\Phi_{\tau}\}_{\tau
\in\mathbb{R}}$ of canonical transformations of $T^{\ast}\left(
Q\times\mathbb{R}\right)  $ generated by $\widetilde{X}_{\widetilde{H}}$
\ will permute the $(\widetilde{\Sigma}_{0})_{t}$'s \ among themselves, i.e:%
\begin{equation}
\Phi_{\tau}:(\widetilde{\Sigma}_{0})_{t}\rightarrow(\widetilde{\Sigma}%
_{0})_{t+\tau};\text{ }\tau\in\mathbb{R};\text{ }\Phi_{\tau}\circ\Phi
_{\tau^{\prime}}=\Phi_{\tau+\tau^{\prime}};\text{ }\Phi_{0}=Id_{\widetilde
{\mathcal{M}}}\label{evol}%
\end{equation}

\begin{remark}
One should keep in mind that, while on the extended phase space $T^{\ast
}\left(  Q\times\mathbb{R}\right)  $, due to the fact that \ the dynamical
vector field $\widetilde{X}_{\widetilde{H}}$ does not depend explicitly on the
evolution parameter $\tau$, the dynamics is "autonomous" \cite{AM,MSSV}, and
gives rise therefore to a "bona fide" one-parameter group, this is not so on
$T^{\ast}Q$. There, due to the time-dependence, the dynamics will give rise in
general \cite{AM,MMN} to a \textbf{two}-parameter family $\{\phi_{t_{2}t_{1}%
}\}$ of canonical transformations of $T^{\ast}Q$ obeying the law of
combination \cite{MMN}:%
\begin{equation}
\phi_{t_{3}t_{2}}\circ\phi_{t_{2}t_{1}}=\phi_{t_{3}t_{1}};\text{ }\phi
_{t_{1}t_{1}}=Id_{T^{\ast}Q}%
\end{equation}
which will "collapse", of course, into a one-parameter group if the
Hamiltonian happens to be time-independent, i.e., in such a case: $\phi
_{t_{2}t_{1}}=\phi_{t_{2}-t_{1}}$.Explicitly (see also Eqs.(\ref{can4}) and
(\ref{can5})), the one-parameter group $\{\Phi_{\tau}\}$ will act as
\cite{MMN}:%
\begin{equation}
\Phi_{\tau}:\left(  m,t,h\right)  \rightarrow\left(  \phi_{t+\tau,t}\left(
m\right)  ,t+\tau,h+H\left(  m,t\right)  -H\left(  \phi_{t+\tau}\left(
m\right)  ,t+\tau\right)  \right)  \label{evol2}%
\end{equation}

\end{remark}

At this point, it becomes apparent how one can cast the time-dependent $HJ$
problem as stated in Eq.(\ref{time_dep}) into a geometrical form on the
extended phase space $\widetilde{\mathcal{M}}$, paralleling to some extent the
discussion in Sect.\ref{TIHJ2}. \ With any $S\in\mathcal{F}(\widetilde{Q})$
($dS\in\mathfrak{X}^{\ast}(\widetilde{Q})$) we $\ $can associate the map:%
\begin{equation}
\widetilde{\varphi}_{S}:\widetilde{Q}\rightarrow\widetilde{\mathcal{M}};\text{
}\widetilde{\varphi}_{S}:\widetilde{Q}\ni\left(  q,t\right)  \rightarrow
\left(  q,\frac{\partial S}{\partial q},t,\frac{\partial S}{\partial
t}\right)  \subset\widetilde{\mathcal{M}}%
\end{equation}
which is a global section w.r.t. the projection map: $\widetilde{\pi
}:\widetilde{\mathcal{M}}\rightarrow\widetilde{Q}$. Then:
\begin{equation}
\widetilde{\Gamma}=:\widetilde{\varphi}_{S}(\widetilde{Q})\subset
\widetilde{\mathcal{M}}%
\end{equation}
will be the graph of the closed (actually exact) one-form $dS$ on
$\widetilde{Q}$, and hence a ($\left(  n+1\right)  $-dimensional)
transversal Lagrangian submanifold in $\widetilde{\mathcal{M}}$. The
requirement that $S$ be a solution of the $PDE$ (\ref{time_dep}) can
be rephrased in
geometrical terms by requiring that:%
\begin{equation}
\widetilde{\varphi}_{S}^{\ast}\widetilde{H}=0\Longleftrightarrow
\widetilde{\Gamma}\subset\widetilde{\Sigma}_{0}\label{TDHJ1}%
\end{equation}

\begin{remark}
Eq.(\ref{TDHJ1}) should make it clear why, at variance with the
time-independent case, the "zero-level set"  $\widetilde{\Sigma}_{0}$ plays a
distinguished r\^{o}le in the geometrical approach to the time-dependent $HJ$ problem.
\end{remark}

\bigskip

The geometrical meaning of the time-dependent $HJ$ problem in the present
formulation is therefore the following:

\textit{We look for transversal Lagrangian submanifolds }$\widetilde{\Gamma}$
\textit{in }$\widetilde{\mathcal{M}}$\textit{ that are contained in
}$\widetilde{\Sigma}_{0}$\textit{ and are graphs of exact one-forms on
}$\widetilde{Q}$\textit{. }$\widetilde{\Gamma}$ \textit{will have codimension
}$n$ \textit{in \ }$\widetilde{\Sigma}_{0}$. \textit{A complete solution of
the time-dependent }$HJ$ \textit{problem will be an }$n$-\textit{dimensional
foliation of the "zero-level set" }$\widetilde{\Sigma}_{0}$ \textit{by such
submanifolds. Usually, these submanifolds are defined by the "dispersion relations" of our $PDE$. }

By reasoning as in Sect.\ref{TIHJ2} we can conclude that the vector field
$\widetilde{X}_{\widetilde{H}}$ \ will be tangent to the submanifold
$\widetilde{\Gamma}$. Hence, the dynamical flow will leave it \ invariant,
i.e.:%
\begin{equation}
\Phi_{\tau}(\widetilde{\Gamma})=\widetilde{\Gamma}\text{ }\forall
\tau\label{evol3}%
\end{equation}

All of the above picture is on the extended phase space $\widetilde
{\mathcal{M}}$, and one should see now how it can be made to "descend" to the
physical phase space $\mathcal{M}=T^{\ast}Q$, i.e. how it behaves under the
projection:
\begin{equation}
\pi^{\prime}:T^{\ast}\left(  Q\times\mathbb{R}\right)  =\widetilde
{\mathcal{M}}\rightarrow\mathcal{M=}T^{\ast}Q;\text{ }\pi^{\prime}:\left(
m,t,h\right)  \rightarrow m\label{projection}%
\end{equation}

For every fixed $t$ we may consider $S$ as a function $S_{t}\in\mathcal{F}%
\left(  Q\right)  $ with $dS_{t}\in\mathfrak{X}^{\ast}\left(  Q\right)  $.
Then, the map:%
\begin{equation}
\varphi_{S,t}:Q\ni q\rightarrow\left(  q,\left(  dS_{t}\right)  \left(
q\right)  \right)  \in T^{\ast}Q
\end{equation}
giving a global section w.r.t. the projection: $\pi_{0}:T^{\ast}Q\rightarrow
Q$, will define the transversal Lagrangian submanifold $\Gamma_{t}%
=\varphi_{S,t}\left(  Q\right)  $ in $T^{\ast}Q$, the graph of the exact
one-form $dS_{t}$.

Going back now to the situation in \ $\widetilde{\mathcal{M}}=T^{\ast}\left(
Q\times\mathbb{R}\right)  $, we can consider the family of intersections
$\{\widetilde{\Gamma}\cap(\widetilde{\Sigma}_{0})_{t}\}_{T\in\mathbb{R}}$ of
$\widetilde{\Gamma}$ with the constant-$t$ sections of $\widetilde{\Sigma}%
_{0}$. Explicitly:%
\begin{equation}
\widetilde{\Gamma}\cap(\widetilde{\Sigma}_{0})_{t}=\left\{  \left(
q,\frac{\partial S}{\text{ }\partial q},t,\frac{\partial S}{\partial
t}\right)  \text{ }\forall t\text{ }\right\}
\end{equation}
and, for every $t$, $\widetilde{\Gamma}\cap(\widetilde{\Sigma}_{0})_{t}$ will
be an $n$-dimensional (hence isotropic \cite{MMN} in $\widetilde{\mathcal{M}}%
$) submanifold contained in $(\widetilde{\Sigma}_{0})_{t}$. Using then
Eqs.(\ref{evol}) and (\ref{evol3}) one sees that $\Phi_{\tau}$ permutes again
these submanifolds among themselves, i.e.:%
\begin{equation}
\Phi_{\tau}\left(  \widetilde{\Gamma}\cap(\widetilde{\Sigma}_{0})_{t}\right)
=\widetilde{\Gamma}\cap(\widetilde{\Sigma}_{0})_{t+\tau}\label{evol4}%
\end{equation}
and, under the projection (\ref{projection}) we obtain:%
\begin{equation}
\pi^{\prime}\left(  \widetilde{\Gamma}\cap(\widetilde{\Sigma}_{0})_{t}\right)
=\Gamma_{t}%
\end{equation}
Moreover, using Eq.(\ref{evol2}), we see that the $\Gamma_{t}$'s evolve in
time as:%
\begin{equation}
\phi_{t+\tau,t}\left(  \Gamma_{t}\right)  =\Gamma_{t+\tau};\text{ }t,\tau
\in\mathbb{R}\label{evol5}%
\end{equation}

We obtain then the following picture:

\textit{Any solution of the }$HJ$ \textit{problem in }$\widetilde{\mathcal{M}%
}=T^{\ast}\left(  Q\times\mathbb{R}\right)  $, \textit{i.e. any single
geometrical object }$\widetilde{\Gamma}$, \textit{gives rise in }%
$\mathcal{M}=T^{\ast}Q$ \textit{to a family }$\left\{  \Gamma_{t}\right\}  $
\textit{of time-dependent, Lagrangian submanifolds in }$T^{\ast}Q$,
\textit{each }$\Gamma_{t}$ \textit{being the graph of an exact one-form
}$dS_{t}$ on $Q$, \textit{that evolves in time according to Eq.(\ref{evol5}).}

One should keep in mind, however, that time-evolution need not be, and quite
often is not,  a harmless process, to the extent that solutions that are "well
behaved" at a certain time may develop caustics and/or become ill-behaved at
later times. To "cure" these and other possible pathologies, and also to fully
implement symmetries in the time-dependent context as well as in the
time-independent one (see also the discussion at the end of
Subsect.\ref{TIHJ2}), one might be forced to  require  that the relevant
submanifolds be graphs of closed but not necessarily exact one-forms and/or to
abandon the requirement of transversality. We will not discuss here these
generalizations, but  refer rather to the literature \cite{MMN} for further details.

\subsection{The Hamilton-Jacobi Problem on Lie Groups.}

In the previous Subsection we have seen how, even for
non-conservative systems, one can implement canonically, via the
one-parameter group $(\Phi_{\tau })_{\tau\in\mathbb{R}}$, the action
of the Abelian group $\mathbb{R}$ of time translations by suitably
enlarging the phase space from $T^{\ast}Q$ to $T^{\ast}\left(
Q\times\mathbb{R}\right)  $. It is interesting to generalize this
approach \cite{MMN}to the case in which the action of the Abelian
group $\mathbb{R}$ is replaced by that of a more general Lie group
$\mathbb{G}$.
\subsubsection{Canonical Actions of Lie Groups on Symplectic Manifolds.}\label{can_act1}
Let $\mathcal{M}=T^{\ast}Q$ with the canonical symplectic
form\footnote{Although we will concentrate here on $T^{\ast}Q$, most
of what will be said applies equally well \cite{MMN} to a general symplectic
manifold $\left( \mathcal{M},\omega\right)  $.} $\omega_{0}$ and let
$\mathfrak{g}$ be the Lie algebra of a $k$-dimensional Lie group $\
\mathbb{G}$, with a basis $\left\{
e_{r}\right\}  _{r=1}^{k}$ obeying the Lie bracket relations:%
\begin{equation}
\left[  e_{r},e_{s}\right]  =C_{rs}^{t}e_{t}%
\end{equation}
the $C_{rs}^{t}$'s being the structure constants of the group.

The group $\mathbb{G}$ will act canonically \cite{AM,MMN,MSSV} on
$T^{\ast}Q$ if there exists a realization $\left\{  \phi_{g}\right\}
_{g\in\mathbb{G}}$
via a family of diffeomorphisms of $T^{\ast}Q$ satisfying:%
\begin{equation}
\phi_{g}\circ\phi_{g^{\prime}}=\phi_{gg^{\prime}\text{ }}\forall
g,g^{\prime
}\in\mathbb{G};\text{ }\phi_{e}=Id_{T^{\ast}Q}%
\end{equation}
(with $e$ the identity of the group) and:%
\begin{equation}
\phi_{g}^{\ast}\omega_{0}=\omega_{0}\text{ }\forall g\in\mathbb{G}
\label{canon}%
\end{equation}

To the basis $\left\{  e_{r}\right\}  _{r=1}^{k}$ of the Lie algebra
there will correspond a set $\left\{  X_{r}\right\}
_{r=1}^{k},X_{r}\in \mathfrak{X}\left(  T^{\ast}Q\right) ,r=1,...,k$
of fundamental \cite{MSSV} vector fields generating the finite
(canonical) transformations $\phi_{g}$ and obeying the commutation
relations:
\begin{equation}
\left[  X_{r},X_{s}\right]  =C_{rs}^{t}X_{t} \label{fund1}%
\end{equation}

As a consequence of Eq.(\ref{canon}) the $X_{r}$'s will be locally
Hamiltonian, i.e.: $L_{X_{r}}\omega_{0}=0$ $\forall r$. We shall assume
them to be actually globally Hamiltonian, i.e. that there exist
globally defined
functions $H_{1},...,H_{k}$ s.t.:%
\begin{equation}
i_{X_{r}}\omega_{0}=-dH_{r},\text{ }r=1,...,k \label{fund2}%
\end{equation}

The Poisson brackets of the $H$'s will be given as usual, by:
$\left\{ H_{r},H_{s}\right\}  =i_{X_{s}}i_{X_{r}}\omega_{0}$, and it
is a simple matter to show that Eqs.(\ref{fund1}) and (\ref{fund2})
imply:
\begin{equation}
d\left[  \left\{  H_{r},H_{s}\right\}  +C_{rs}^{t}H_{t}\right]  =0
\end{equation}
i.e.:%
\begin{equation}
\left\{  H_{r},H_{s}\right\}  =-C_{rs}^{t}H_{t}+d_{rs};\text{ \ }d_{rs}%
\in\mathbb{R},\text{ }d_{rs}+d_{sr}=0 \label{Poisson5}%
\end{equation}
The Jacobi identity on the Poisson bracket (\ref{Poisson5}) implies then:%

\begin{equation}
C_{rs}^{u}d_{ut}+\text{ }cycl.\text{ }perm.\text{ }of\text{ }\left(
r,s,t\right)  =0 \label{Jacobi6}%
\end{equation}
One can get rid of the additional constants if there exist other
constants
$\lambda_{1},...,\lambda_{k}$ s.t.:%
\begin{equation}
C_{rs}^{t}\lambda_{t}=d_{rs} \label{Jacobi7}%
\end{equation}
in which case, re-defining: $\widetilde{H}_{r}=:H_{r}-\lambda_{r}$
one gets
Poisson brackets without additional additive constants, i.e.:%
\begin{equation}
\left\{  \widetilde{H}_{r},\widetilde{H}_{s}\right\}
=C_{rs}^{t}\widetilde
{H}_{t}%
\end{equation}
\

\begin{remark}
The condition (\ref{Jacobi6}) is the statement that the map: $\textbf{d}:\mathfrak{g}%
\times\mathfrak{g}\rightarrow\mathbb{R}$ \ acting via: $\textbf{d}\left(  e_{r}%
,e_{s}\right)  =:d_{rs}$ (with $\mathfrak{g}$ acting trivially on $\mathbb{R}%
$) must be a two-cocycle \cite{CE,MM,MMR,MMSS} in $\mathfrak{g}$, while
Eq.(\ref{Jacobi7}) requires it to be also a coboundary. This will be
granted if the second cohomology group $H^{2}\left(
\mathfrak{g},\mathbb{R}\right)$  of $\mathfrak{g}$ with values
in $\mathbb{R}$ \cite{CE} \ vanishes, i.e.: $H^{2}\left(
\mathfrak{g},\mathbb{R}\right)  =0$. This is true for many of the groups that are relevant for Classical Dynamics \cite{SM}as well as for Quantum Mechanics and Field Theory such as the Euclidean, the Lorentz and the Poincar$\acute{e}$ groups. It fails however to be true for the Galilei group, which has a non-trivial two-cocycle (i.e. a cocycle which is not a coboundary) connected with the mass \cite{SM} which is  an invariant for the Galilei group.
\end{remark}
\subsubsection{The Canonical Action of a Lie group $\mathbb{G}$ on $T^{\ast}\mathbb{G}$.}\label{can_act2}
As is well known \cite{MSSV}, a Lie group $\mathbb{G}$ can act on
itself, among other ways, via left or right translations. We will
concentrate on the former, i.e. on the
diffeomorphisms:%
\begin{equation}
L_{g}^{\left(  0\right)  }:g^{\prime}\rightarrow L_{g}^{\left(
0\right)
}g^{\prime}=:gg^{\prime};\text{ }L_{g_{1}}^{\left(  0\right)  }\circ L_{g_{2}%
}^{\left(  0\right)  }=L_{g_{1}g_{2}}^{\left(  0\right)  };\text{
}\forall
g,g^{\prime},g_{1},g_{2}\in\mathbb{G} \label{left-trans}%
\end{equation}
and will discuss briefly how this action can be lifted to a
canonical action of $\mathbb{G}$ on the cotangent bundle
$T^{\ast}\mathbb{G}$.

Left translations are generated by a set: $X_{1}^{\left(  0\right)
},...,X_{k}^{\left(  0\right)  }$ of independent and nowhere
vanishing
right-invariant \cite{MSSV} vector fields satisfying:%
\begin{equation}
\left[  X_{r}^{\left(  0\right)  },X_{s}^{\left(  0\right)  }\right]
=C_{rs}^{t}X_{t}^{\left(  0\right)  } \label{left_inv}%
\end{equation}
with associated dual (right-invariant) one-forms $\theta^{\left(
0\right)
r},r=1,...,k$ such that:%
\begin{equation}
i_{X_{s}^{\left(  0\right)  }}\theta^{\left(  0\right)
r}=\delta_{s}^{r}
\label{dual}%
\end{equation}
and satisfying the Maurer-Cartan structure equations \cite{MSSV}:%
\begin{equation}
d\theta^{\left(  0\right)  r}+\frac{1}{2}C_{st}^{r}\theta^{\left(
0\right)
s}\wedge\theta^{\left(  0\right)  t}=0 \label{Maurer}%
\end{equation}
These right-invariant one-forms will provide a basis $\theta^{\left(
0\right)  r}\left(  g\right)  $ for the cotangent space $T_{g}^{\ast
}\mathbb{G}$ at any point $g\in\mathbb{G}$.

A basis for the dual $\mathfrak{g}^{\ast}$ of the Lie algebra
$\mathfrak{g}$ will be denoted as $\left\{  e^{r}\right\}
_{r=1}^{k}$ and will be specified
by:%
\begin{equation}
\left\langle e^{r}|e_{s}\right\rangle =\delta_{s}^{r}%
\end{equation}
Any element $h\in T_{g}^{\ast}\mathbb{G}$ can be written as: $h=h_{r}%
\theta^{\left(  0\right)  r}\left(  g\right)  $, and we have the association:%
\begin{equation}
T_{g}^{\ast}\mathbb{G}\ni h=h_{r}\theta^{\left(  0\right)  r}\left(
g\right)
\leftrightarrow\widetilde{h}=h_{r}e^{r}\in\mathfrak{g}^{\ast}%
\end{equation}
This implies \cite{MSSV} that $T^{\ast}\mathbb{G}$ will be
diffeomorphic to
$\mathbb{G}\times\mathfrak{g}^{\ast}$: $T^{\ast}\mathbb{G\simeq G}%
\times\mathfrak{g}^{\ast}$.

Using the canonical projection:
$\pi:T^{\ast}\mathbb{G}\rightarrow\mathbb{G}$ we can pull back the
$\theta^{\left(  0\right)  r}$'s to obtain: $\theta
^{r}=:\pi^{\ast}\theta^{\left(  0\right)
r}\in\mathfrak{X}^{\ast}\left( T^{\ast}\mathbb{G}\right)  ,$
$r=1,...,k$, and we can define then the
canonical one-form $\Theta_{0}$ on $T^{\ast}\mathbb{G}$ as:%
\begin{equation}
\Theta_{0}=h_{r}\theta^{r}\label{can7}%
\end{equation}
Hence: $\Omega_{0}=d\Theta_{0}$ will be the canonical symplectic
form on $T^{\ast}\mathbb{G}$.

The canonical lifts $X_{r}^{\left(  0\right)  \ast}$ from
$\mathbb{G}$ to $T^{\ast}\mathbb{G}$ of the $X_{r}^{\left(  0\right)
}$\ 's will be defined
\cite{MSSV} by the conditions:%
\begin{equation}
L_{X_{r}^{\left(  0\right)  \ast}}\Theta_{0}=0,\text{ }T\pi\left(
X_{r}^{\left(  0\right)  \ast}\right)  =X_{r}^{\left(  0\right)
}\,\ r=1,...,k \label{can-lift1}%
\end{equation}
i.e. $X_{r}^{\left(  0\right)  \ast}$ must be a vector field that
leaves the canonical one-form unchanged and that projects down under
the tangent map $T\pi$ to $X_{r}^{\left(  0\right)  }$ for all
$r$'s. $X_{r}^{\left( 0\right)  \ast}$ is of course Hamiltonian as
the first of Eqs.(\ref{can-lift1})
is equivalent to: $i_{X_{r}^{\left(  0\right)  \ast}}\Omega_{0}=-d(i_{X_{r}%
^{\left(  0\right)  \ast}}\Theta_{0}).$ The $X_{r}^{\left(  0\right)
\ast}$'s
turn out also \cite{MMN} to be dual to the $\theta^{r}$'s. i.e.:%
\begin{equation}
i_{X_{r}^{\left(  0\right)
\ast}}\theta^{s}=\delta_{r}^{s}\Longrightarrow
i_{X_{r}^{\left(  0\right)  \ast}}\Theta_{0}=h_{r}%
\end{equation}
and hence:%
\begin{equation}
i_{X_{r}^{\left(  0\right)  \ast}}\Omega_{0}=-dh_{r}%
\end{equation}

The $X_{r}^{\left(  0\right)  \ast}$'s close on the same Lie algebra
as the $X_{r}^{\left(  0\right)  }$'s \cite{SM}, i.e.:
$[X_{r}^{\left(  0\right) \ast},X_{s}^{\left(  0\right)
\ast}]=C_{rs}^{t}X_{t}^{\left(  0\right)  \ast }$. The same happens
with the Poisson brackets of the $h_{r}$'s, but in this case (cfr. Eq.(\ref{Poisson5}))
without \cite{MSSV,SM} additional constants i.e.:
\begin{equation}
\left\{  h_{r},h_{s}\right\}  =-C_{rs}^{t}h_{t}%
\end{equation}
with no need to require here the second cohomology group $H^{2}(\mathfrak{g}%
,\mathbb{R)}$ to vanish.

It is the lifted vector fields $X_{r}^{\left(  0\right)  \ast}$ that
generate the canonical action of $\mathbb{G}$ on
$T^{\ast}\mathbb{G}$. Finite transformations will be denoted as
$T^{\ast}L_{g}^{\left(  0\right)  }$ and
they will act as \cite{MMN,SM}:%
\begin{equation}
T^{\ast}L_{g}^{\left(  0\right)  }:\left(
g^{\prime},\widetilde{h}^{\prime }\right)  \longrightarrow\left(
gg^{\prime},\mathcal{D}\left(  g\right)
\widetilde{h}^{\prime}\right)
\end{equation}
where $\mathcal{D}\left(  g\right)  $ is the coadjoint \cite{MSSV}
representation of $\mathbb{G}$ on $\mathfrak{g}^{\ast}$.

\subsubsection{The Hamilton-Jacobi Problem on $\mathbb{G}$.}
We shall extend now our formulation of the $TDHJ$ problem by replacing $\mathbb{R}$ with $\mathbb{G}$.
We can combine the canonical actions of $\mathbb{G}$ \ on $T^{\ast}Q$ and
on $T^{\ast}\mathbb{G}\simeq\mathbb{G\times}\mathfrak{g}^{\ast}$ to get a
canonical action on: $\widetilde{\mathcal{M}}=:$ $T^{\ast}\left(
Q\times\mathbb{G}\right)  $.  $\widetilde{\mathcal{M}}$ will be \ a symplectic
manfold with the canonical two-form:%
\begin{equation}
\omega=\omega_{0}+\Omega_{0}%
\end{equation}
Points in $\widetilde{\mathcal{M}}$ will be denoted as: $\widetilde
{m}=(m,g,\widetilde{h}),m\in T^{\ast}Q,g\in\mathbb{G}$ and $\widetilde{h}%
\in\mathfrak{g}^{\ast}$. The required canonical action will be denoted as
$\Phi_{g},g\in\mathbb{G}$, and will be given by:%
\begin{equation}
\Phi_{g}:\left(  m,g^{\prime},\widetilde{h}^{\prime}\right)  =\left(  \phi
_{g}\left(  m\right)  ,gg^{\prime},\mathcal{D}\left(  g\right)  \widetilde
{h}^{\prime}\right)
\end{equation}
It will be generated by the vector fields:%
\begin{equation}
\widetilde{X}_{r}=X_{r}+X_{r}^{\left(  0\right)  \ast}, r=1,...,k%
\end{equation}
which will obey the commutation relations:%
\begin{equation}
\left[  \widetilde{X}_{r},\widetilde{X}_{s}\right]  =C_{rs}^{t}\widetilde
{X}_{t}%
\end{equation}
and will be Hamiltonian w.r.t. the symplectic form $\omega$ with Hamiltonians:%
\begin{equation}
\widetilde{H}_{r}=H_{r}+h_{r}%
\end{equation}

One can prove, just as in Subsect.\ref{can_act1}, that the Poisson brackets of
the $\widetilde{H}_{r}$'s are given by:%
\begin{equation}
\widetilde{\left\{  \widetilde{H}_{r},\widetilde{H}_{s}\right\}  }=-C_{rs}%
^{t}\widetilde{H}_{t}+d_{rs}%
\end{equation}
and they will "inherit" the same two-cocycle $d$ that we found to be present  there.

We can define now the momentum map \cite{MSSV} $\mu$ as:%
\begin{equation}
\mu:\widetilde{\mathcal{M}}\longrightarrow\mathfrak{g}^{\ast}:\mu\left(
\widetilde{m}\right)  =\widetilde{H}_{r}\left(  \widetilde{m}\right)
e^{r}=\left(  H_{r}\left(  m\right)  +h_{r}\right)  e^{r}%
\end{equation}
and define next the analog $\widetilde{\Sigma}$ of the submanifold
$\widetilde{\Sigma}_{0\text{ }}$of Subsect.\ref{TDHJ_sec} \ will be defined as
the zero-level set of the momentum map, i.e.:%
\begin{equation}
\widetilde{\Sigma}=\mu^{-1}\left(  0\right)  =\left\{  \left(  m,g,\widetilde
{h}\right)  |m\in T^{\ast}Q,g\in\mathbb{G},h_{r}=-H_{r}\left(  m\right)
\Leftrightarrow\widetilde{H}_{r}=0\right\}  \subset\widetilde{\mathcal{M}}%
\end{equation}
which is a submanifold of dimension $\left(  2n+k\right)  $ diffeomorphic to
$T^{\ast}Q\times\mathbb{G}$.

Now, as:%
\begin{equation}
\widetilde{\left\{  \widetilde{H}_{r},\widetilde{H}_{s}\right\}  }\equiv
L_{\widetilde{X}_{s}}\widetilde{H}_{r}=-C_{rs}^{t}\widetilde{H}_{t}+d_{rs}%
\end{equation}
we see that $\widetilde{\Sigma}$ will be invariant under the canonical action
$\Phi_{g}$ if and only if $\textbf{d}=0$. This is one of the most compelling reasons
for requiring $\textbf{d}$ to be a coboundary, i.e.\ for requiring that $H^{2}\left(
\mathfrak{g,}\mathbb{R}\right)  =0$.

In analogy with the constant-$t$ sections introduced in Subsect.\ref{TDHJ_sec}
we can introduce "constant-$g$ sections" $\widetilde{\Sigma}_{g}$:%
\begin{equation}
\widetilde{\Sigma}_{g}=\left\{  (m,g,\widetilde{h})|m\in T^{\ast}Q,\text{
}g\text{ }fixed,\text{ }\widetilde{H}_{r}=0\right\}
\end{equation}
which are of dimension $2n$ and diffeomorphic to $T^{\ast}Q$. The canonical
action $\Phi_{g}$ permutes these sections among themselves, i.e.:%
\begin{equation}
\Phi_{g}\left(  \widetilde{\Sigma}_{g^{\prime}}\right)  =\widetilde{\Sigma
}_{gg^{\prime}}%
\end{equation}

One can pose now a geometrical  $HJ$ problem for the Lie group $\mathbb{G}$ as follows:

\textit{Find a (maximal) Lagrangian (hence of dimension }$\left(  n+k\right)
$\textit{) submanifold }$\widetilde{\Gamma}\subset\widetilde{\mathcal{M}%
}=T^{\ast}\left(  Q\times\mathbb{G}\right)  $, \textit{transversal w.r.t. the
projection: }$\widetilde{\pi}:\widetilde{\mathcal{M}}\rightarrow\mathcal{M}$
\textit{and } \textit{such that: }%
\begin{equation}
i_{\widetilde{\Gamma}}^{\ast}\mu=0\Leftrightarrow i_{\widetilde{\Gamma}}%
^{\ast}\widetilde{H}_{r}=0,\text{ }r=1,..,k\Rightarrow\widetilde{\Gamma
}\subset\widetilde{\Sigma}\label{trans-Lag}%
\end{equation}
\textit{and sucht that, for every }$g\in\mathbb{G}$\textit{:}%
\begin{equation}
\Gamma_{g}=\pi\left(  \widetilde{\Gamma}\cap\widetilde{\Sigma}_{g}\right)
=:\Gamma_{g}%
\end{equation}
\textit{is a }$g$\textit{-dependent Lagrangian submanifold in }$\mathcal{M}%
=T^{\ast}Q$.

This implies, of course:%
\begin{equation}
\Phi_{g}\left(  \widetilde{\Gamma}\right)  =\widetilde{\Gamma}%
\end{equation}
as well as, at the level of $\mathcal{M}$:%
\begin{equation}
\phi_{g}\left(  \Gamma_{g^{\prime}}\right)  =\Gamma_{gg^{\prime}}%
\end{equation}

\begin{remark}
Requiring the simultaneous vanishing of all the $\widetilde{H}_{r}$'s on
$\widetilde{\Gamma}$ is reminiscent of the joint $HJ$ problems that were
discussed in Sect.\ref{joint0}. The vanishing of the two-cocycle $\textbf{d}$ is then
another necessary condition for the $HJ$ problem on Lie groups to admit of a solution.
\end{remark}

Finally, if $\widetilde{\Gamma}$ happens to be the graph of an exact one-form
$dS\in\mathfrak{X}\left(  Q\times\mathbb{G}\right)  $ determined by
$S\in\mathcal{F}\left(  Q\times\mathbb{G}\right)  $, we can express
everything in somewhat more familiar terms.

In the notations of Refs.\cite{MMN,SM}, let: $\alpha=\left(  \alpha
^{1},...,\alpha^{k}\right)  $ be local coordinates for $\mathbb{G}$ in a
neighborhood of the identity, normalized to $\alpha^{r}=0,r=1,...,k$, at the
identity. Then, if $\alpha,\beta,\gamma$ are coordinates for $a,b,c\in
\mathbb{G}$, the composition law: $ab=c$ is expressed in local coordinates as:%
\begin{equation}
\gamma^{r}=f^{r}\left(  \alpha,\beta\right)  ,\text{ }r=1,...,k
\end{equation}
with preassigned functions\footnote{The conditions to which the $f^{r}$'s have
to obey are discussed at length in Ref.\cite{SM}.} $f^{r}$. Introducing the
(invertible \cite{SM}) matrix:%
\begin{equation}
\eta\left(  \alpha\right)  =\left\Vert \eta_{s}^{r}\left(  \alpha\right)
\right\Vert ;\eta_{s}^{r}\left(  \alpha\right)  =\frac{\partial f^{r}\left(
\beta,\alpha\right)  }{\partial\beta^{s}}|_{\beta=0}\text{ }%
\end{equation}
and its inverse: $\xi\left(  \alpha\right)  =\eta\left(  \alpha\right)  ^{-1}$
($\eta_{s}^{r}\left(  0\right)  =\xi_{s}^{r}\left(  0\right)  =\delta_{s}^{r}%
$), right-invariant vector fields are given \cite{MMN}by:%
\begin{equation}
X_{r}^{\left(  0\right)  }\left(  \alpha\right)  =-\eta_{r}^{s}\left(
\alpha\right)  \frac{\partial}{\partial\alpha^{s}}%
\end{equation}
and the associated right-invariant one-forms will be given by:%
\begin{equation}
\theta^{\left(  0\right)  r}=-\xi_{s}^{r}d\alpha^{s}%
\end{equation}
The canonical one-form $\Theta_{0}$ of Eq.(\ref{can7}) can then be rewritten
as:%
\begin{equation}
\Theta_{0}=-h_{s}\xi_{r}^{s}d\alpha^{r}=:\pi_{r}d\alpha^{r}%
\end{equation}
and will define the "momenta":%
\begin{equation}
\pi_{r}=-\xi_{r}^{s}h_{s}\Leftrightarrow h_{r}=-\eta_{r}^{s}\pi_{s}%
\end{equation}

Then, with: $S=S\left(  q;\alpha\right)  $ and the usual replacement:
$\pi\rightarrow\partial S/\partial\alpha$, the conditions: $\widetilde{H}%
_{r}=H_{r}+h_{r}=0$ become:%
\begin{equation}
H_{r}\left(  q;\frac{\partial S\left(  q;\alpha\right)  }{\partial q}\right)
-\eta_{r}^{s}\left(  \alpha\right)  \frac{\partial S\left(  q;\alpha\right)
}{\partial\alpha^{s}}=0
\end{equation}
and this is now the (conventional) system of $PDE$'s associated with a
canonical realization of the Lie group $\mathbb{G}$ on $T^{\ast}Q$.

For further generalizations we refer once again to Ref.\cite{MMN}.

\section{Concluding Remarks.}

Following Dirac's prescription \cite{Dir3} according to which Classical Mechanics must be
a suitable limit of Quantum Mechanics, we have considered in this paper the
Hamilton-Jacobi theory as emerging from Quantum Mechanics when we consider an
approximation in which Planck's constant is treated as a parameter.

>From this point of view, while we keep on with the originating idea
of the theory of considering Hamiltonian Optics as a suitable limit
of Wave Optics, we have taken advantage of the fact that Wave
Mechanics concerns also with particles with internal structure. This
has suggested that we deal not only with "scalar differential
operators" like Schr\"{o}dinger's or Klein-Gordon's, but also with
"matrix-valued differential operators" as those appearing in the
Pauli as well as in the Dirac equations. The net result is an
extension of the usual Hamilton-Jacobi formalism to a formalism
where the Hamiltonian "scalar function" is replaced by a
matrix-valued Hamiltonian.

In the same spirit, following what happens in relativistic field theories, we
have replaced the one-parameter group of time evolution with a Lie group, e.g.
the Poincar\'{e} group.

We have not considered the Hamilton-Jacobi theory for field theories
proper but, following the ideas outlined in this paper, it should
not be very difficult to foresee how to proceed.

The joint Hamilton-Jacobi problem turns out also to be very useful
to deal with holonomic and non-holonomic constraints within the
Hamilton-Jacobi formalism \cite{CGMM2}. In this connection we should
also mention some recent relevant contributions \cite{ILM,LMM} to
the same subject.

The geometrization of the Hamilton-Jacobi problem presented here has
many advantages over more conventional presentations. For
example, as mentioned already in Sect.\ref{TIHJ2}, posing a "Hamilton-Jacobi problem" as the search of foliations of the energy surfaces by Lagrangian but not necessarily transversal submanifolds opens the possibility of implementing the full set of canonical symmetries as symmetries of the Hamilton-Jacobi problem as well. We have also shown elsewhere \cite{MMN} how to deal, in the same fully geometric spirit, with transformations and symmetries for partial differential equations of the Monge-Amp\`{e}re type. Using the present
generalization to differential operators acting on sections of
vector bundles it should be possible to incorporate into the
formalism more general $PDE$'s than those of the above
"Monge-Amp\`{e}re" type.

Using our matrix-valued Hamiltonians it will be possible to deal
with equations of the Wong type \cite{BMSS1}, i.e. equations
describing particles interacting with Yang-Mills fields. Also, in
this approach, treating, say, electrons moving in some monopole-like
magnetic field, algebroids arise in quite a natural way. We shall
postpone more details on these aspects to a forthcoming paper.





\subsection*{Acknowledgments}
The first two authors would like to thank for hospitality, with the
support of the INFN-MEC joint exchange program, the Departamento de
Mat\'{e}maticas, Universidad Carlos III (Madrid) where part of this
work was brought to completion.



\medskip
Received xxxx 20xx; revised xxxx 20xx.
\medskip

\end{document}